\documentclass[11pt,a4paper]{article}
\usepackage{hyperref}
\usepackage{epsfig}
\usepackage{amsmath}
\usepackage{amssymb}
\usepackage{color}
\usepackage[footnotesize]{caption}
\usepackage[subrefformat=parens,labelformat=parens]{subfig}
\usepackage{cite}
\newcommand{\eVq}  {\text{eV}^2}

\numberwithin{equation}{section}

\begin{document}

\vskip 1cm

\begin{center}
{\Large\bf Nonzero $\theta_{13}$ and Leptogenesis in a Type-I See-saw Model
with $A_4$ Symmetry} 
%\\[2mm]
\vskip 2cm

{ Biswajit Karmakar$^{a,}$\footnote{k.biswajit@iitg.ernet.in}, Arunansu
Sil$^{a,}$\footnote{asil@iitg.ernet.in}}\\[3mm]
{\it{
$^a$ Indian Institute of Technology Guwahati, 781039 Assam, India}
}
\end{center}

\vskip 1cm

\begin{abstract}
\noindent 
In the light of recent discovery of nonzero $\theta_{13}$, we have analyzed the 
Altarelli-Feruglio $A_4$ flavor symmetry model extended with additional flavon. 
The inclusion of the new field leads to the deviation from exact tri-bimaximal 
neutrino mixing pattern in the context of type-I see-saw by producing a nonzero 
$\theta_{13}$ consistent with the recent experimental results at the leading 
order. A sum rule for light neutrino masses is also obtained in this context.   
The set-up constraints the two Majorana phases involved in the lepton mixing 
matrix in terms of $A_4$ parameter space. We have shown that a nonzero
lepton asymmetry  can be generated while next-to-leading order contributions to 
the neutrino Yukawa couplings  are considered. The two Majorana phases play 
crucial role in CP-asymmetry parameter and the involvement of $\theta_{13}$ 
in it, is exercised. 
\end{abstract}

\thispagestyle{empty}

\newpage

%%%%%%%%%%%%%%%%%%%%%%%%%%%%%%%%%%%%%%%%%%%%%%%
%%%%%%%%%%%%%%				 %%%%%%%%%%%%%%%%%%%%%%%
%%%%%%%%%%%%%	         Introduction    %%%%%%%%%%%%%%%%%%%%%%%
%%%%%%%%%%%%%%				  %%%%%%%%%%%%%%%%%%%%%%
%%%%%%%%%%%%%%%%%%%%%%%%%%%%%%%%%%%%%%%%%%%%%%%

\section{Introduction}\label{sec1}

The evidence of non-vanishing value of the mixing angle $\theta_{13}$ from 
several experiments (Double Chooz\cite{Abe:2011fz}, Daya Bay\cite{DayaBay},  
RENO \cite{Ahn:2012nd}, T2K \cite{Abe:2013hdq}), receives particular attention
in these days since the precise determination of neutrino mixing would be
crucial for better understanding the issues related to the flavor. In this
context it is important to study the neutrino mass matrix, $m_{\nu}$, that can
be structured from discrete flavor symmetry. The neutrino mass matrix 
$m_{\nu}$,
in general,  can be diagonalized by the $U_{PMNS}$ matrix (in the basis where
charged leptons are diagonal) as 
\begin{equation}
m_{\nu} = U^*_{PMNS} diag(m_1, m_2, m_3)U^{\dagger}_{PMNS},
\end{equation}
where $m_1, m_2, m_3$ are the real mass eigenvalues. The standard 
parametrization \cite{Beringer:1900zz} of the $U_{PMNS}$ matrix is given by 
{\small 
\begin{eqnarray}
U_{PMNS}=\begin{bmatrix} %\nonumber
    c_{12}c_{13}                                &   s_{12}c_{13}                
               & s_{13}e^{-i\delta}\\
    -s_{12}c_{23}-c_{12}s_{13}s_{23}e^{i\delta} &
c_{12}c_{23}-s_{12}s_{13}s_{23}e^{i\delta}    &   c_{13}s_{23} \\
     s_{12}s_{23}-c_{12}s_{13}c_{23}e^{i\delta} &
-c_{12}s_{23}-s_{12}s_{13}c_{23}e^{i\delta}    &   c_{13}c_{23} 
\end{bmatrix}
\begin{bmatrix}
  1 &0 &0\\
  0 &e^{i\alpha_{21}/2} &0\\
  0 &0 &e^{i\alpha_{31}/2}
\end{bmatrix},
\label{upmns}
\end{eqnarray}
}
where $c_{ij}=\cos\theta_{ij}$, $s_{ij}=\sin\theta_{ij}$, $\delta$ is the 
CP-violating Dirac phase while $\alpha_{21}$ and $\alpha_{31}$ are the two
CP-violating Majorana phases. Though the neutrino mixing angles $\theta_{12}$,
$\theta_{23}$ and the two mass-squared differences have been well measured at
several neutrino oscillation experiments \cite{NuExpt}, only an upper bound was
present (consistent with zero) for the other mixing angle $\theta_{13}$ till
2011 \cite{Schwetz:2008er}. Then the recent results from Double Chooz 
\cite{Abe:2011fz}, Daya Bay \cite{DayaBay}, RENO \cite{Ahn:2012nd}, T2K 
\cite{Abe:2013hdq}, suggest that in fact $\theta_{13}$ is nonzero and of 
sizable magnitude. From the updated global analysis \cite{Forero:2014bxa} 
involving all the data from neutrino experiments, we have 1$\sigma$ and 
3$\sigma$ ranges of mixing angles and the mass-squared differences as mentioned 
(NH and IH stand for the normal and inverted mass hierarchies respectively) in 
Table \ref{table1}. Majorana phases are not appearing in neutrino oscillation 
probability and therefore can not be constrained from neutrino oscillation data 
directly \cite{ExcptMajo}. As of now, any specific constraint on the Dirac CP 
violating phase $\delta$ is still missing and so it is ranged between 0 to 
2$\pi$ \cite{Forero:2014bxa}.
\begin{table}[h]\centering
\resizebox{12.9cm}{!}{
  \begin{tabular}{|c|c|c|c|}
    \hline
    Oscillation parameters & $1\sigma$ range&  3$\sigma$ range
    \\
    \hline\hline
    $\Delta m^2_{21}$ 
   & 7.42--7.79 [$10^{-5}\hspace{.1cm} \eVq$]  & 7.11--8.18 \\
   \hline
    $|\Delta m^2_{31}|$
    &
    \begin{tabular}{c}
      $2.41-2.53$ [$10^{-3}\hspace{.1cm}\eVq$] (NH)\\
      $2.32-2.43$ [$10^{-3}\hspace{.1cm}\eVq$] (IH)\\
    \end{tabular}
    &
    \begin{tabular}{c}
      $2.30-2.65$\\
      $2.20-2.54$
    \end{tabular}
    \\
    \hline
    $\sin^2\theta_{12}$
    & 0.307--0.339 & 0.278--0.375\\
    \hline
    $\sin^2\theta_{23}$
    &
    \begin{tabular}{c}
      0.439--0.599 (NH)\\
      0.530--0.598 (IH)\\
    \end{tabular}
    &
    \begin{tabular}{c}
      0.392--0.643\\ 
      0.403--0.640 
    \end{tabular} \\
    \hline
    $\sin^2\theta_{13}$
    &
    \begin{tabular}{c}
      0.0214--0.0254 (NH) \\
      0.0221--0.0259 (IH)
    \end{tabular}
    &
   \begin{tabular}{c}
      0.0177--0.0294 \\
      0.0183--0.0297
    \end{tabular} \\
    \hline
       \hline
     \end{tabular}
}
     \caption{ \label{table1} {\small Summary of neutrino oscillation parameters
     for normal and inverted neutrino mass hierarchy from the analysis of
     \cite{Forero:2014bxa}.}}
\end{table}

This clearly indicates a completely different pattern of mixing in the 
lepton sector compared to the quark sector. Efforts therefore have been 
exercised for a long time in realizing the neutrino mixing pattern and among 
them patterns based on discrete flavor groups attract particular attention. A 
case of special mention is where
$\sin^2\theta_{12} = 1/3$, $\sin^2\theta_{23} = 1/2$ along with $\sin\theta_{13}
=0$ resulted,  called the tri-bimaximal (TBM) mixing pattern 
\cite{Harrison:1999cf}. Note that all these mixing angles inclusive of 
vanishing $\theta_{13}$ were in the right ballpark of experimental findings 
before 2011. Many discrete groups have been employed \cite{King:2013eh} in 
realizing the TBM 
mixing pattern, and $A_4$ turned out to be a special one which can reproduce 
this pattern in a most economic way 
\cite{Ma:2004zv,Altarelli:2005yp,Altarelli:2005yx}. 
$A_4$ is a discrete group of even permutations of four objects. It has three
 inequivalent one-dimensional representations ($1,1^{\prime},1^{\prime\prime}$)
and a three dimensional representation (3). In this work, we mostly concentrate
on Altarelli-Feruglio (AF) type of model \cite{Altarelli:2005yx} where the
light neutrino masses are generated through type-I see-saw mechanism. So the
right handed neutrinos ($N^c$) are introduced which transform as a triplet of
$A_4$. Flavon fields transforming trivially and non-trivially under the $A_4$
are also introduced, whose vacuum expectation values (VEV) break the $A_4$
flavor symmetry at some high scale. The framework is supersymmetric and based 
on the Standard Model (SM) gauge interactions. As it was argued in 
\cite{Altarelli:2005yx}, the introduction of supersymmetry was instrumental 
to provide the correct vacuum alignment. Then the type-I see-saw leads to the 
TBM mixing in the light neutrinos while the charged lepton mass matrix is found 
to be diagonal. 

However with the latest developments toward the nonzero value of
$\theta_{13}$, it is essential to modify the exact TBM pattern. Several attempts
were made in this direction during last couple of years in the context of
$A_4$-based flavor models \cite{Nonzerotheta1,Brahmachari:2008fn,
Nonzerotheta2,Shimizu:2011xg,Ma:2011yi,King:2011zj,Nonzerotheta3,
Altarelli:2012bn , Nonzerotheta4}. It is to be noted from these analysis that
inclusion of higher order terms only would not produce a sufficiently large 
$\theta_{13}$ as predicted by experiments. So a leading order deformation of 
the original $A_4$ model is required which we will study in this work.
%A detailed classicification of models based on $A_{4}$ has been presented 
%in ref.\cite{Ding:2011gt}. 

Another important phenomenon that can not be realized in the context of the 
Standard Model is to explain the observed matter-antimatter asymmetry of the
Universe. However it is known that the standard weak interactions can lead to 
processes (mediated by sphalerons) which can convert the baryons and leptons. 
So a baryon asymmetry can be effectively generated from a lepton asymmetry. The 
mechanism for generating the lepton asymmetry is called leptogenesis 
\cite{Davidson:2008bu}. The discussion of it is of particular importance here, 
while explaining the generation of light neutrino mass through type-I see-saw 
mechanism. The inclusion of heavy right handed (RH) neutrinos in the framework 
provides the opportunity to discuss also the leptogenesis scenario through the 
CP-violating decay of it in the early Universe. Although the ingredients (RH 
neutrinos) are present, it is known that the see-saw models predicting the 
exact TBM structure can not generate the required lepton asymmetry 
\cite{Jenkins:2008rb}, the reason being the term involved in the asymmetry 
related to the neutrino Yuwaka coupling matrix is proportional to the 
identity matrix and thus the lepton number asymmetry parameter vanishes. 
However it was shown in \cite{Jenkins:2008rb} that one can in principle 
consider higher dimensional operators in the neutrino Yukawa couplings of the 
model. The effect of this inclusion is to deviate the products of the 
Yukawa-terms in lepton asymmetry parameter from unity and thereby generating 
 nonzero lepton asymmetry. 

In this work, our aim is to produce nonzero $\theta_{13}$ as well as to realize
leptogenesis in the same framework. We have extended the flavon-sector of AF
\cite{Altarelli:2005yx} by introducing an extra flavon, $\xi'$ which transforms
as $1'$ under $A_4$. Similar sort of extensions have been considered in 
\cite{Brahmachari:2008fn,Shimizu:2011xg}. However the analyses in those works 
are mostly related to the deviation over the final form of $m_{\nu}$ obtained 
from AF model, while 
here we consider modification of $m_{\nu}$ through the deviation from the 
RH neutrino mass matrix $M_R$. In \cite{King:2011zj,Cooper:2011rh}, a 
perturbative deviation 
from tri-bimaximal mixing is considered through $M_{R}$, though leptogenesis 
was not considered in that framework. This provides the opportunity to 
analyze $M_R$ in detail and the effect on the Majorana phases can also be 
studied. Inclusion of $Z_3$ symmetry in the model forbids several
unwanted terms and thus helps in constructing specific structure of the coupling
matrices. While the charged lepton mass matrix is found to be in the diagonal
form, the RH neutrino mass matrix has an additional structure originated from
$\xi'$-related term. Due to this, the light neutrino diagonalizing matrix no
longer remains in TBM form rather a deviation is resulted which leads to
nonzero $\theta_{13}$. In the RH neutrino mass matrix, three complex parameters
$a, b$ and $d$ are present. We found that the low energy observables can be
expressed in terms of two parameters $\lambda_1 (=|d/a|)$, $\lambda_2 (=|b/a|)$;
relative phase between $b$ and $a$ ($\phi_{ba}$) and $|a|$. The relative phase
between $d$ and $a$ are assumed to be zero for simplicity. We have studied
the dependence of $\theta_{13}$ on $\lambda_1$. The allowed range of
$\theta_{13}$ restricts the range of the parameter space of  $\lambda_1$. 
Then following the analysis \cite{Hagedorn:2009jy}, we are able to constrain 
also the Majorana phases ($\alpha_{21}$, $\alpha_{31}$) involved in the 
$U_{PMNS}$ and study their dependence on the parameter $\lambda_2$ (for this we 
have fixed $\lambda_1$ to its value that corresponds to the best-fit value of 
$\sin^2\theta_{13}$) for both normal and inverted hierarchy cases. In this 
scenario, we obtain a general sum rule involving the light neutrino masses 
$m_{i=1,2,3}$ and the Majorana phases, $\alpha_{21}$, $\alpha_{31}$. The 
effective mass parameter involved in the neutrinoless double beta decay is also 
estimated. We then investigate the generation of lepton asymmetry from the 
decay of RH neutrinos within `one flavor approximation'
\cite{Hagedorn:2009jy,Pascoli:2006ie,Blanchet:2006ch,Davidson:2008pf}.
As previously stated, nonzero lepton
asymmetry can be obtained once we include the next to leading order terms in
the Yukawa sector. Note that this inclusion does not spoil the diagonal nature
of charged lepton mass matrix. The explicit appearance of these Majorana phases
in the CP-asymmetry parameter, $\epsilon_i$, provides the possibility of 
studying the dependence of $\epsilon_i$ on $\lambda_2$. The expression of 
$\epsilon_i$ also involves the $\theta_{13}$ mixing angle in our set-up. Since 
$\theta_{13}$ depends on $\lambda_1$, we have also studied the variation of 
$\epsilon_i$ (or baryon asymmetry $Y_B$) against $\theta_{13}$ while 
$\lambda_2$ is fixed at a suitable value. 

In section \ref{sec2}, we describe the structure of the model by specifying 
the fields involved and their transformation properties under the symmetries 
imposed. Then in section {\ref{sec3}}, we discuss the eigenvalues and phases 
involved in the RH neutrino sector. We also find the lepton mixing matrix and study the 
correlation between the mixing angles in terms of $\lambda_1$. Section 
\ref{sec4} is devoted to study the Majorana phases, light neutrino masses, 
effective mass parameter involved in neutrinoless double beta decay. 
Leptogenesis is analysed in section \ref{lep} and following that, we have 
conclusion in section \ref{conc}.

%%%%%%%%%%%%%%%%%%%%%%%%%%%%%%%%%%%%%%%%%%%%%%%
%%%%%%%%%%%%%%				 %%%%%%%%%%%%%%%%%%%%%%%
%%%%%%%%%%%%%	         Section 2       %%%%%%%%%%%%%%%%%%%%%%%
%%%%%%%%%%%%%%				  %%%%%%%%%%%%%%%%%%%%%%
%%%%%%%%%%%%%%%%%%%%%%%%%%%%%%%%%%%%%%%%%%%%%%%

\section{Structure of The Model}\label{sec2}
%%%% Symmetries and Matter Content%%%%%%%%%%
We consider here an extension of the original Altarelli-Feruglio(AF) model
\cite{Altarelli:2005yx} (with right-handed neutrinos) for generating lepton
masses and mixing by introducing one additional flavon $\xi'$ which transforms 
as $1'$ under $A_4$. We will find this modification turns out to be 
instrumental to have nonzero $\theta_{13}$. The particle content and the 
symmetries of the model are provided in Table \ref{t2}. The framework is 
supersymmetric and the gauge group is same as that of the Standard Model. All 
the left handed doublets $L_{i(=1,2,3)}$ transform as $A_4$ triplets, and the 
RH charged leptons $e^c,\mu^c,\tau^c$ are $A_4$ singlets $1$, $1''$, $1'$ 
respectively. In order to realize the type-I see-saw, three right handed 
neutrinos ($N_i^c$) are considered which are triplets of $A_4$. The flavor 
symmetry $A_4$ is accompanied by a discrete $Z_3$ symmetry, which forbids 
several unwanted terms. The $A_4$ multiplication rules are mentioned in 
appendix \ref{apa}. There are four flavons ($\phi_S$, $\phi_T$, $\xi$, $\xi'$) 
in the model, which are SM gauge singlets.  When the flavons (the scalar 
component of it) get vacuum expectation values (vev), 
$\langle\phi_{S}\rangle=(v_{S},v_{S}, v_{S})$,\ 
$ \langle\phi_{T}\rangle=(v_{T},0,0)$,\ $ \langle\xi\rangle=u$,\ 
$\langle\xi'\rangle=u'$, the $A_4 \times Z_3$ symmetry is broken and 
generates the flavor structure of the sector. The fields $\phi^S_0, \phi^T_0$ 
and $ \xi_0 $ are the driving fields, carrying two units of $U(1)_R$ 
charges, introduced to realize the vacuum alignments of the flavon fields, 
$\phi_S, \phi_T, \xi, \xi'$. Supersymmetry helps in realizing this vacuum 
alignment by setting the F-term to be zero. A brief discussion on the 
vacuum alignment is provided in appendix \ref{apb}. $H_u$ and $H_d$ are 
the two Higgs doublets present in the set-up transforming as singlets under
$A_4$ with the vevs $v_u$ and $v_d$ respectively.  
\begin{table}[h]
\centering
\resizebox{12.9cm}{!}{%
\begin{tabular}{|l|ccccccccccc||ccc||}
\hline
 & $e^{c}$ & $\mu^{c}$ & $\tau^{c}$  & $L_i$ & $N_i^{c}$ & $H_{u}$ &$H_{d}$ & 
$\phi_{S}$ & $\phi_{T}$ &  $\xi$ & $\xi'$ &
 $\phi_{0}^{S}$ & $\phi_{0}^{T}$ & $\xi_{0}$\\
\hline
$A_{4}$ & 1 & $1''$ & $1'$ &  3 & 3 & 1 & 1 & 3 & 3 & 1&$1'$& 3 & 3& 1\\
\hline
$Z_{3}$ &$\omega$ & $\omega$ & $\omega$ & $\omega$ & $\omega^{2}$ & 1 &
$\omega$ 
& $\omega^{2}$ & 1 & $\omega^{2}$ & $\omega^{2}$
& $\omega^2$ & 1 & $\omega^{2}$ \\
\hline
$U(1)_{R}$ & 1 & 1 &  1 & 1 & 1 & 0 & 0 & 0 & 0 & 0 & 0 & 2 & 2 & 2\\
\hline
\end{tabular}
}\
\caption{\label{t2} {\small Fields content and transformation properties under
the symmetries imposed on the model. Here $\omega$ is the third root of
unity.}}
\end{table}
With the above mentioned field configuration, the effective superpotential for 
the charged lepton sector contains the following terms in the leading order
(LO), 

\begin{equation}\label{cl}
 w_{L}= \Big[{y}_{e}e^c(L{\phi_{T}}) +{y}_{\mu}\mu^c(L{\phi_{T}})'+ 
{y}_{\tau}\tau^c(L{\phi_{T}})''\Big]
 \left(\frac{H_{d}}{\Lambda}\right),\\%\nonumber
\end{equation}
where $\Lambda$ is the cut-off scale of the theory and ${y}_{e}$, ${y}_{\mu}$,
${y}_{\tau}$ are the coupling constants. Terms in the first parenthesis 
represent products of two triplets (here $L$ and ${\phi_{T}}$ for example) 
under $A_4$, each of these terms contracts with $A_4$ singlets $1$, $1''$ and 
$1'$ corresponding to $e^c$, $\mu^c$ and $\tau^c$ respectively. Finally it
sets the charged lepton coupling matrix as the diagonal one in the leading 
order,  
\begin{eqnarray}
Y_{L} = \frac{v_{T}}{\Lambda}  \begin{bmatrix} %\nonumber
      y_{e} &0 &0\\
       0 &y_{\mu} &0\\
       0 &0 &y_{\tau}
\end{bmatrix},
\end{eqnarray}
\noindent
once the flavon vevs as well as the Higgs vevs are inserted. The relative 
hierarchies between the charged leptons can be generated if one introduces 
global Froggatt-Nielsen ($U(1)_{FN}$) flavor symmetry, under which RH charged 
leptons have different charges in addition to a FN field 
\cite{Froggatt:1978nt,Altarelli:2010gt}.

In absence of the $\xi'$ field, the neutrino sector would have the
superpotential of the form 
\begin{equation}\label{wnu}
 {{w}}_{\nu}=y(N^cL)H_{u} + x_{A}\xi(N^cN^c)+x_{B}(N^cN^c\phi_{S}),\\
\end{equation}
which yields the Dirac ($m_D$) and Majorana ($M_R$) neutrino mass matrices at 
the LO as given by
\begin{eqnarray}
m_{D} =yv_u \begin{bmatrix} %\nonumber
      1 &0 &0\\
       0 &0 &1\\
       0 &1 &0
\end{bmatrix}\equiv Y_{\nu{0}}v_u {\text {;}} \hspace{.3cm}
M_{R} = \begin{bmatrix}
      a+2b/3   &-b/3    &-b/3\\
       -b/3    &2b/3    &a-b/3\\
       -b/3    &a-b/3   &2b/3
\end{bmatrix},\label{Mr0}
\end{eqnarray}
where $a = 2x_Au, b = 2x_B v_S$ and $Y_{\nu{0}}$ can be taken as the LO neutrino
Yukawa coupling matrix. Here  $y$, $x_{A}$, and $x_{B}$ are respective 
coupling constants. It has been known \cite{Ma:2004zv,Altarelli:2005yp, 
Altarelli:2005yx} that this kind of structure produces the exact TBM mixing, 
predicting $\theta_{13}=0$. However in our setup, the inclusion of $\xi'$ 
ensures the presence of another term in the superpotential $w_{\nu}$, given by
\begin{equation}
 x_{N}\xi'(N^cN^c),
\end{equation}
at the LO, where $x_N$ is another coupling constant. It introduces a modified
Majorana mass matrix, compared to the one ($M_R$) in TBM case, having the form 
\begin{eqnarray}\label{Mrd}
M_{Rd} = \begin{bmatrix}
      a+2b/3 &-b/3 &-b/3\\
       -b/3 &2b/3 &a-b/3\\
       -b/3 &a-b/3 &2b/3
\end{bmatrix}+
\begin{bmatrix}
      0 & 0 & d\\
      0 & d & 0\\
      d & 0 & 0
\end{bmatrix},
\end{eqnarray}
where $d=2x_{N}u'$. Since this additional term is also at the renormalizable 
level, we expect the term $d$ to be of the order of $a$ and $b$, in general. 
Inclusion of higher order terms in $m_{D}$ would be very important in having 
leptogenesis as we will discuss it in section \ref{lep}. 

%%%%%%%%%% NLO contributions %%%%%%%%%%%%%%%
In general we expect the vevs of the flavon fields ($v_S$, $v_T$, $u$, $u'$) 
are of same order of magnitude $\sim$ $v$ (say). Therefore, the magnitude of  
light neutrino $m_{\nu}$ becomes $\sim(yv_u)^2/v$, generated through
type-I see-saw mechanism. However there could be operators like $(LH_u)(LH_u)$,
which can also contribute to the light neutrino mass. In our model such terms
appear only in combination with $\phi_S$, $\xi$ and $\xi'$ in quadrature 
$\left(LH_uLH_u\frac{1}{\Lambda^3}\left[\phi_S^2,\phi_S\xi,\phi_S\xi', 
\xi\xi',\xi'^2\right]\right)$, as $LH_uLH_u$ is not an invariant under $Z_3$. 
Note that these terms contribute to the light neutrino mass of order 
$\frac{v_u^2}{v}\kappa^3$ where $\kappa=\frac{v}{\Lambda}\ll1$. Hence 
they are relatively small compared to the neutrino mass generated from
type-I see-saw by order of $\kappa^3$ with $y\sim\mathcal{O}(1)$ or so and 
therefore can be neglected in the subsequent analysis. 

There are next-to-leading order (NLO) corrections present in the model which 
are suppressed by $1/\Lambda^n$ with $n\geq1$. For the charged lepton, the 
leading order (LO) contribution $f^c(L\phi_T)\frac{H_{d}}{\Lambda}$ 
($f^c=e^c,\tau^c, \mu^c$), is already $1/\Lambda$ suppressed. So possible NLO 
contributions are $f^c(L(\phi_T\phi_T)_A)\frac{H_{d}}{\Lambda^2}$ and  
$f^c(L(\phi_T\phi_T)_S)\frac{H_{d}}{\Lambda^2}$, where the suffixes $A$ and $S$
stand for anti-symmetric and symmetric triplet components from the product of
two triplets in the first parenthesis under $A_4$. Now the first term
essentially vanishes from the direction of vevs of $\phi_T$ and the contribution
 coming from the second term is again diagonal, similar to the one obtained 
from LO term. So a mere redefinition of $y_e, y_\mu, y_\nu$ would keep the 
charged lepton matrix as a diagonal, even if NLO contributions are 
incorporated. This conclusion is in line of earlier observation
\cite{Altarelli:2005yx,Jenkins:2008rb}. 

We could as well include higher order terms involving $1/\Lambda$ (which are 
allowed by all the symmetries imposed) to the neutrino Yukawa coupling  as 
$x_{C}(N^{c}L)_S\phi_{T}H_{u}/\Lambda + x_{D}(N^{c}L)_A\phi_{T}H_{u}/\Lambda$,
with $x_{C}$ and $x_{D}$ as coupling constants. Therefore, at the 
next-to-leading order, the neutrino Yukawa coupling matrix can be re-written
as, 
\begin{eqnarray}
Y_{\nu} & = & Y_{{\nu}0}+\delta{Y_{\nu}}\nonumber\\
        & = & y  \begin{bmatrix}\label{dynu}
      1 &0 &0\\
       0 &0 &1\\
       0 &1 &0
\end{bmatrix}+\frac{x_{C}v_{T}}{\Lambda}\begin{bmatrix}
              2 & 0 & 0\\
              0 & 0 & -1\\
              0 & -1 & 0
              \end{bmatrix}+\frac{x_{D}v_{T}}{\Lambda}\begin{bmatrix}
              0 & 0 & 0\\
              0 & 0 & -1\\
              0 & 1& 0
              \end{bmatrix}.
\end{eqnarray}
\noindent
%The same type of $\delta{Y_{\nu}}$ was also obtained in \cite{Jenkins:2008rb, 
%Hagedorn:2009jy}. 
This will not produce any significant effect on the 
light neutrino masses and mixing obtained through type-I see-saw mechanism
primarily with leading order $m_D$ and $M_{Rd}$, as those terms are suppressed
by the cut-off scale $\Lambda$ compared to the LO contribution. However
these will have important role in leptogenesis, what we will discuss in section 
\ref{lep}. 

For RH Majorana neutrinos, the non-vanishing NLO corrections in the mass matrix 
arise from the following terms
\begin{eqnarray}
 \delta{M_{Rd}}&=&C_1(N^cN^c)_S\phi_T\xi/\Lambda+C_2(N^cN^c)_A\phi_T\xi'/\Lambda
               +C_3(N^cN^c)(\phi_S\phi_T)/\Lambda\nonumber\\
               &+&C_4(N^cN^c)^{''}(\phi_S\phi_T)^{'}/\Lambda
               +C_5(N^cN^c)^{'}(\phi_S\phi_T)^{''}/\Lambda\nonumber\\
               &+&C_6(N^cN^c)_{S}(\phi_S\phi_T)_{S}/\Lambda
               +C_7(N^cN^c)_{S}(\phi_S\phi_T)_{A} /\Lambda.
\end{eqnarray}
Here $C_{i=1,..,7}$ are the respective couplings and prefixes $'$ and $''$
correspond to the $1'$ and $1''$ singlets of $A_4$ produced from the
multiplication of two triplets under $A_4$ within $(...)$. Terms proportional to
$C_3$ and $C_4$ can be absorbed in $M_{Rd}$ and contributions from the 
remaining terms produce a deviation from $M_{Rd}$ that can be written in a 
compact form as
\begin{eqnarray}
\Delta{M_{Rd}}=\begin{bmatrix} \nonumber
              2X_D & X_B  & -X_A\\
              X_B  & 2X_A & X_D\\
             -X_A  & X_D  & X_B
\end{bmatrix},
\end{eqnarray}
\noindent
where $X_D=(3C_6v_s+C_7v_s+C_1u)\kappa$, $X_B=C_5v_s\kappa$ and
$X_A=(2C_7v_s+C_2u_N)\kappa$. Almost similar type of conclusion was 
obtained in \cite{Hagedorn:2009jy}, apart from the fact that we have absorbed
the term proportional to $C_4$ in LO contribution of $M_{Rd}$ and a new 
contribution coming from $C_2$ (through $\xi'$) is included in the 
definition of $X_A$. 

%%%%%%%%%%%%%%%%%%%%%%%%%%%%%%%%%%%%%%%%%%%%%%%
%%%%%%%%%%%%%%				 %%%%%%%%%%%%%%%%%%%%%%%
%%%%%%%%%%%%%	         Section 3       %%%%%%%%%%%%%%%%%%%%%%%
%%%%%%%%%%%%%%				  %%%%%%%%%%%%%%%%%%%%%%
%%%%%%%%%%%%%%%%%%%%%%%%%%%%%%%%%%%%%%%%%%%%%%%

\section{Neutrino Masses and Mixing} \label{sec3}

Light neutrino mass matrix is obtained through the type-I see-saw mechanism 
 as $m_{\nu}=m_{D}^{T}M^{-1}m_{D}$, where $M$ is 
the Majorana mass matrix for RH neutrinos. Note that the Majorana mass matrix 
$M$, with the form $M_R$ as in Eq.(\ref{Mr0}) (${\it i.e.}$ without  $\xi'$
field), can be diagonalized through $U^T_{TB} M_R U_{TB} = {\rm{diag}} 
(M_{R1}e^{i\phi_1}, M_{R2}e^{i\phi_2}, M_{R3}e^{i\phi_3})$, where $U_{TB}$
exhibits the TBM mixing pattern \cite{Harrison:1999cf} and is described by, 
\begin{eqnarray}\label{utb}
       U_{TB}&=& \begin{bmatrix}
      \sqrt{\frac{2}{3}} &\frac{1}{\sqrt{3}} &0\\
      -\frac{1}{\sqrt{6}} &\frac{1}{\sqrt{3}} &\frac{1}{\sqrt{2}}\\
      -\frac{1}{\sqrt{6}} &\frac{1}{\sqrt{3}} &-\frac{1}{\sqrt{2}}
\end{bmatrix},
\end{eqnarray}
and $M_{R1,2,3}$ are  given by $|b+a|$, $|a|$ and $|b-a|$ respectively.
$\phi_{1,2,3}$ are the arguments of the eigenvalues respectively. 
It is found \cite{King:2011zj} that the light neutrino mass matrix $m_{\nu}$ 
($= m^T_D M^{-1}_R m_D$) in this case can also be diagonalized by a matrix $U$
which is same as $U_{TB}$ except the second and third rows of it are
interchanged (apart from the phases involved), so as to have $U^T m_{\nu} U =
{\rm diag} (m_1, m_2, m_3)$. The light neutrino mass eigenvalues $m_i$ are given
by $m_i = \left( yv_u \right)^2/M_{Ri}$, and they can be made real and 
positive since the phase of $y$ can be absorbed due to redefinition of phases 
in lepton doublets and the phases $\phi_i$ can be
included in the diagonal phase matrix of $U$. As previously discussed, this 
structure of $M_R$ is not useful in explaining the nonzero $\theta_{13}$, as 
seen while comparing the above form of $U$ and $U_{PMNS}$. Since the measured 
value of $\theta_{13}$ is not very small, it is difficult to reconcile 
$\theta_{13}$ just by deforming $m_{\nu}$ from its above form with the 
introduction of small expansion parameter \cite{softbrk}. Rather we 
should have deformation parameter at the same order of the existing elements in 
$m_{\nu}$. In our framework, we have introduced the $\xi'$ field for this 
purpose. 

\subsection{RH Neutrinos}
The new scalar singlet $\xi'$ contributes to the heavy RH neutrino sector 
through the $x_{N}\xi'(N^cN^c)$ term and the Majorana neutrino mass matrix then 
takes the form of $M_{Rd}$ as in Eq.(\ref{Mrd}). We note that after having a 
rotation by $U_{TB}$, the $M_{Rd}$ takes the form as given by, 
\begin{eqnarray}
U_{TB}^{T}M_{Rd}U_{TB} & = & \begin{bmatrix}
                             a+b-\frac{d}{2}      &  0  &-\frac{\sqrt{3}}{2}d\\
                             0             & a+d &     0             \\
                              -\frac{\sqrt{3}}{2}d &  0  &   -a+b+\frac{d}{2}
                              \end{bmatrix}.
\end{eqnarray}
Therefore a further rotation by $U_1$ (another unitary matrix) takes the matrix 
$M_{Rd}$ to a diagonal one,  
${\rm diag}(M_{1}e^{i\varphi_{1}},M_{2}e^{i\varphi_{2}},M_{3}e^{i\varphi_{3}}) 
=(U_{TB}U_{1})^{T}M_{Rd}U_{TB}U_{1}$, where $M_{i=1,2,3}$ are given by, 
\begin{equation}
 M_{1}=|b+\sqrt{a^2+d^2-ad}|=|a|\left|\lambda_2e^{i\phi_{ba}}+
\sqrt{1+\lambda^{2}_{1}e^{2i\phi_{da}}-\lambda_1e^{i\phi_{da}}}\right|,\label
{M1}\\
\end{equation}
\begin{equation}
M_{2} =|a+d|=|a|\left|1+\lambda_1e^{i\phi_{da}}\right|,
\label{M2}
\end{equation}
\begin{equation}
M_{3}=|b-\sqrt{a^2+d^2-ad}|=|a|\left|\lambda_2e^{i\phi_{ba}}-
\sqrt{1+\lambda^{2}_{1}e^{2i\phi_{da}}-\lambda_1e^{i\phi_{da}}}\right|,
\label{M3}
\end{equation} 
with $\lambda_1=|d/a|$ and $\lambda_2=|b/a|$. $\phi_{da}=\phi_{d}-\phi_{a}$ and
$\phi_{ba}=\phi_{b}-\phi_{a}$ are the phase differences between $(d, a)$ and
$(b, a)$ respectively. Phases associated with the above masses can be written as
\begin{eqnarray}
 \varphi_{1}&=&{\rm arg}(b+\sqrt{a^{2}+d^{2}-ad}),\label{varphi1}\\
 \varphi_{2}&=&{\rm arg}(a+d),\label{varphi2}\\
 \varphi_{3}&=&{\rm arg}(b-\sqrt{a^{2}+d^{2}-ad}).\label{varphi3}
\end{eqnarray}
For simplicity, we will work with $\phi_{da}=0$. Hence above set 
of eigenvalues and phases can be rewritten as 
\begin{eqnarray}
 M_1&=&|a|\left|\lambda_2e^{i\phi_{ba}}+{\rm K}\right| \hspace{1cm} 
\varphi_1={\rm arg}(b+a{\rm K})\label{M1real},\\
 M_2&=&|a|\left|1+\lambda_1\right| \hspace{1.9cm} \varphi_2={\rm 
arg}(a+d)\label{M2real},\\
 M_3&=&|a|\left|\lambda_2e^{i\phi_{ba}}-{\rm K}\right| \hspace{1cm} 
\varphi_3={\rm arg}(b-a{\rm K})\label{M3real},
\end{eqnarray}
\noindent
where ${\rm K}=\sqrt{1-\lambda_1+\lambda_1^{2}}$. 

\subsection{Light Neutrino Masses and Mixing Angles}
Light neutrino masses obtained via type-I see-saw mechanism through 
$m_{\nu}=m_{D}^{T}M_{Rd}^{-1}m_{D}$ is now given by 
$m_{D}^{T}U_RU_m^*\left[{\rm diag}\left(M_1,M_2,M_3\right)\right]^{-1}
U_m^*U_R^Tm_D$ where $U_R=U_{TB}U_1$ and 
$U_m={\rm diag}(e^{i\varphi_{1}/2},e^{i\varphi_{2}/2},e^{i\varphi_{3}/2})$.
The special form of $m_D$ (see in Eq.(\ref{Mr0})) suggests that $U_R$, with the
second and third rows interchanged, will be the diagonalizing matrix of the 
light neutrino mass matrix $m_{\nu}$ apart from the diagonal phase matrix. Since
the charged lepton mass matrix is already diagonal, the lepton mixing matrix is
given by \cite{King:2011zj}
\begin{equation}
U_{\nu}=\frac{m_D^T}{yv_u}U_{TB}U_{1}^*{\rm 
diag}(e^{i\varphi_{1}/2},e^{i\varphi_{2}/2},e^{i\varphi_{3}/2}),
\label{upmns2}
\end{equation}
so that $m_{\nu}=U_{\nu}^{*}{\rm diag}(m_i)U_{\nu}^{\dagger}$. Note that, the 
light neutrino masses $m_{1,2,3}$ (real and positive)  are given 
by 
\begin{equation}
 m_{i}=\frac{(yv_{u})^{2}}{M_{i}}, \label{mi}
\end{equation}
where $M_{i=1,2,3}$ are taken from Eq.(\ref{M1real} - \ref{M3real}). We can now 
remove one common
phase by setting $\varphi_1=0$. Hence, the final form of unitary matrix that
diagonalizes $m_{\nu}$ is given by
 \begin{eqnarray}
U_{\nu}&=&\frac{m_D^T}{yv_u}U_{TB}\begin{bmatrix}\label{u1}
       \cos\theta    &0                   &\sin\theta e^{-i\psi } \\
       0     &1  &0\\
       -\sin\theta e^{i\psi }     &0                   &   \cos\theta 
\end{bmatrix}{\rm diag}(1,e^{i\varphi_{2}/2},e^{i\varphi_{3}/2}),\\
        &=&\begin{bmatrix} 
           \sqrt{\frac{2}{3}} \cos\theta                     &   1/\sqrt{3}    
         &
           \sqrt{\frac{2}{3}}\sin\theta e^{-i \psi }\\
            -\frac{\cos\theta}{\sqrt{6}}+\frac{\sin\theta }{\sqrt{2}}e^{i\psi} 
& 1/\sqrt{3} 
            &-\frac{\cos\theta}{\sqrt{2}}-\frac{\sin\theta}{\sqrt{6}}e^{-i \psi 
}\\
                       
-\frac{\cos\theta}{\sqrt{6}}-\frac{\sin\theta}{\sqrt{2}}e^{i \psi }& 
1/\sqrt{3} 
 
              &\frac{\cos\theta}{\sqrt{2}}-\frac{\sin\theta}{\sqrt{6}}e^{-i 
\psi }
\end{bmatrix}.\begin{bmatrix}
       1     &0                   &0\\
       0     &e^{i\varphi_{2}/2}  &0\\
       0     &0                   &e^{i\varphi_{3}/2}
\end{bmatrix},\label{upmns3}
\end{eqnarray}
where we have parametrized the extra $U_{1}$ matrix by $\theta$ and $\psi$ and 
employed Eqs.(\ref{Mr0}) and (\ref{utb}). We identify the Majorana phases as
\begin{eqnarray}
\varphi_2=\alpha_{21}\hspace{.3cm}{\rm and}\hspace{.3cm}
\varphi_3=\alpha_{31}.\label{ap31}
\end{eqnarray}
\noindent
In this type of model, using Eqs.(\ref{M1real} - \ref{M3real}) and Eq.(\ref{mi}) 
we find a general sum rule for light neutrino masses satisfying
\begin{equation}
 \frac{1}{m_1}-\frac{2{\rm K}e^{i\alpha_{21}}}{m_2(1+\lambda_1)}
 =\frac{e^{i\alpha_{31}}}{m_{3}}\label{sumrule}.
\end{equation}
Note that in the limit K$\rightarrow$1 ({\textit {i.e.}} with $\lambda_1=0$),
the sum rule is reduced to the one found in
\cite{Hagedorn:2009jy,Altarelli:2009kr}. The Majorana phases $\alpha_{21}$ and
$\alpha_{31}$ are therefore related to the light neutrino masses. They will play
important role in leptogenesis, which we discuss in section {\ref{lep}}. The 
sum rule may carry important consequence in neutrinoless double beta decay.
A study with different sum rules in this direction can be found in
\cite{sumrules}. 

%%%%%%%%%%%%%%%%% Mixing Angles %%%%%%%%
The charged lepton mass-matrix being diagonal, the above form of $U_{\nu}$ 
leads to (see Eq.{\ref{upmns}})
\begin{eqnarray}\label{ang}
\sin\theta_{13}=\sqrt{\frac{2}{3}}\sin\theta, &\hspace{1cm}& \delta=\psi;\\
\sin^2\theta_{12}=\frac{1}{3(1-\sin^2\theta_{13})} \hspace{.5cm} &{\rm and}&
\hspace{.5cm} \sin^2\theta_{23}=\frac{1}{2}
+\frac{1}{\sqrt{2}}\sin\theta_{13}\cos\delta\label{ang2},
\end{eqnarray}
up to the order $\sin^2\theta_{13}$. The study of these correlations in 
presence of $A_4$ are available in the literature 
\cite{King:2011zj,Altarelli:2012bn,angcorl}. For rest of our analysis 
we will consider 
$\psi=0$. The mixing angle $\theta$ is then given by 
\begin{equation}
\tan2\theta =\frac{{\sqrt3}\lambda_1}{(2-\lambda_1)}. \\
\end{equation}
\begin{figure}[h]
\begin{center}
\includegraphics[scale=0.7]{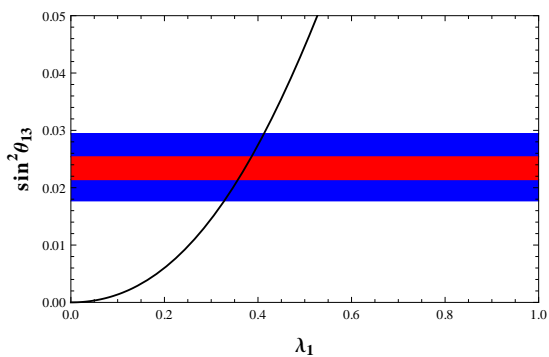}
\caption{{\small $\sin^{2}\theta_{13}$ vs $\lambda_1$ $(i.e.|d/a|)$ plot. 
               Horizontal blue shaded region stands for $3\sigma$ allowed range
               for $\sin^{2}\theta_{13}$ and the red shaded region inside 
               represents $1\sigma$ range for $\sin^{2}\theta_{13}$ obtained
               from \cite{Forero:2014bxa}.}}
\label{fig1}
\end{center}
\end{figure}
\noindent
We have studied the variation of $\sin^2\theta_{13}$ against the parameter 
$\lambda_1$ in Fig.{\ref {fig1}}, where the 1$\sigma$ and 3$\sigma$ allowed 
regions for $\sin^2\theta_{13}$ obtained from \cite{Forero:2014bxa} are also 
indicated in the same by red and blue horizontal shaded regions respectively
for both NH and IH. We observe that for NH, best fit \cite{Forero:2014bxa} value 
of $\sin^2\theta_{13}$ (=0.0234) corresponds to $\lambda_1=0.37$ and that one 
for IH ($\sin^2\theta_{13}$ =0.024) corresponds to $\lambda_1=0.38$.
We also note that the 3$\sigma$ range of $\sin^2\theta_{13}$ covers a narrow
interval of $\lambda_1$ that can be  approximately expressed as  
$0.33\lesssim\lambda_1\lesssim0.41$ as seen from Fig.\ref{fig1} for both NH
and IH. 

\begin{figure}[h]
\centering
%\subfloat[$\sin^{2}\theta_{12}$ vs $\lambda_1$ plot\label{fig2a}]
{
      \includegraphics[scale=0.7]{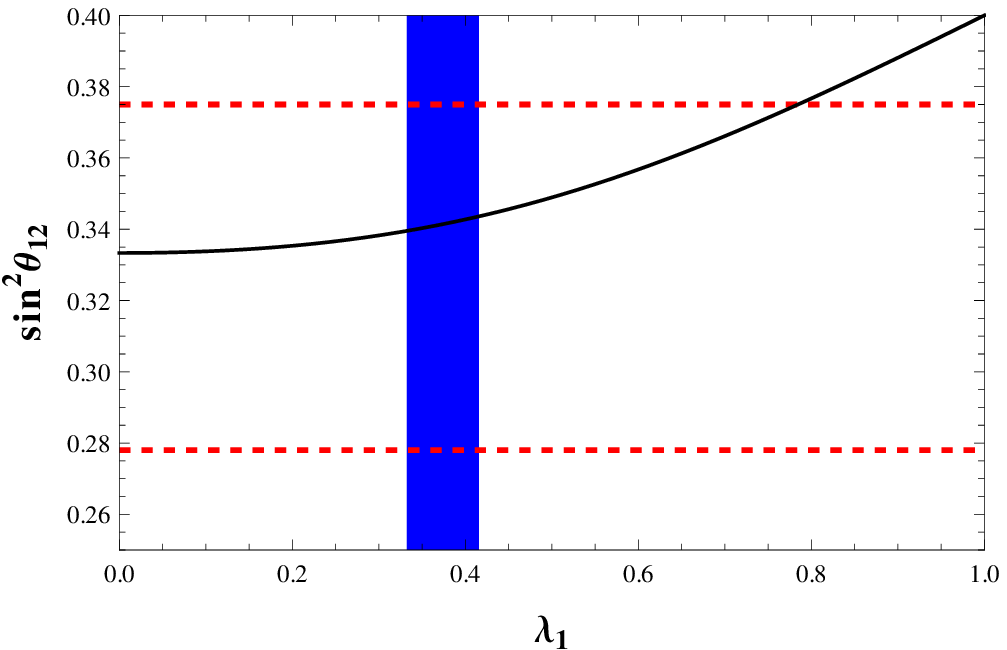}
      }
     % \hfill
     \hspace{.6cm}
  %    \subfloat[$\sin^{2}\theta_{23}$ vs $\lambda_1$ plot\label{fig2b}]
      {
      \includegraphics[scale=0.7]{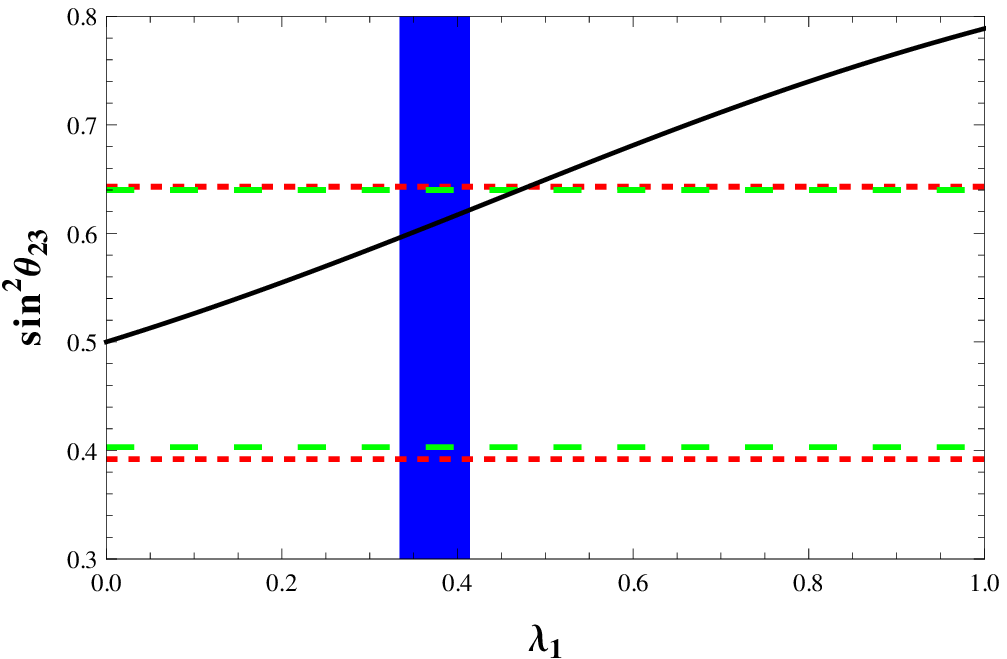}
      }
\caption{{\small  $\lambda_1$ dependence of $\sin^2\theta_{12}$ and 
        $\sin^2\theta_{23}$.
        Vertical blue patch indicates the restricted region of parameter space
        for $\lambda_1$ obtained from Fig.{\ref{fig1}}. Horizontal red dashed 
        lines represent $3\sigma$ allowed range for $\sin^2\theta_{12}$ in the
        left panel, while in the right panel horizontal red dashed and green 
        large-dashed lines represent 3$\sigma$ allowed regions for 
        $\sin^2\theta_{23}$ both NH and IH respectively as in 
\cite{Forero:2014bxa}.}
        }
\label{fig2}
\end{figure}

The other mixing angles $\theta_{12}$ and $\theta_{23}$ are also studied 
through the variation of $\sin^2\theta_{12}$ and $\sin^2\theta_{23}$ against
$\lambda_1$, using Eq.(\ref{ang2}) in Fig.\ref{fig2}. Note that once we
restrict $\lambda_1$ to be in the above mentioned  range (indicated in
Fig.\ref{fig2} by vertical (blue) patches) so that $\sin^2\theta_{13}$ falls 
within the $3\sigma$ allowed range, it constraints the ranges of
$\sin^2\theta_{12}$ and $\sin^2\theta_{23}$ in our set-up. This result is
mentioned in Table {\ref{para2}} as obtained from Fig.{\ref{fig2}}. The 
ranges are well within the $3\sigma$ allowed regions of
$\sin^2\theta_{12}$ and $\sin^2\theta_{23}$ \cite{Forero:2014bxa}.
So we conclude that this particular range of $\lambda_1$
($0.33\lesssim\lambda_1\lesssim0.41$) is consistent in producing all the three
mixing angles successfully, and  we will use this range of $\lambda_1$, while
studying any other observables against $\lambda_1$ unless otherwise stated.

\begin{table}[h]
\centering
\begin{tabular}{|c|c|c|}
\hline
Range of $\lambda_1$ obtained from Fig.{\ref{fig1}}&$\sin 
^{2}\theta_{12}$&$\sin^{2}\theta_{23}$ \\
\hline
 $0.36\lesssim\lambda_1\lesssim0.39$                       & 0.341-0.342   
  &0.604-0.614\\
\hline
$0.33\lesssim\lambda_1\lesssim0.41$                        &  0.339-0.343     &
0.595-0.620 \\
\hline
\end{tabular}
\caption{{\small Allowed regions of $\sin ^{2}\theta_{12}$ and
                 $\sin^{2}\theta_{23}$ obtained from Fig.\ref{fig2} for a
                 restricted range of $\lambda_1$ (corresponding to 
                 Fig.\ref{fig1}) in our set-up.}}\label{para2}
\end{table}

%%%%%%%%%%%%%%%%%%%%%%%%%%%%%%%%%%%%%%%%%%%%%%%
%%%%%%%%%%%%%%				 %%%%%%%%%%%%%%%%%%%%%%%
%%%%%%%%%%%%%	         Section 4        %%%%%%%%%%%%%%%%%%%%%%%
%%%%%%%%%%%%%%				  %%%%%%%%%%%%%%%%%%%%%%
%%%%%%%%%%%%%%%%%%%%%%%%%%%%%%%%%%%%%%%%%%%%%%%

\section{Constraints on parameters from neutrino oscillation
data}\label{sec4}

Apart from $\lambda_1$, we have other parameters $\lambda_2$, $|a|$ and 
$\phi_{ba}$ (after setting $\phi_{da}=0$) in the right handed neutrino sector. 
Note that $\lambda_1$, $\lambda_2$ and $\phi_{ba}$ can be constrained by 
neutrino oscillation data through the ratio of solar and atmospheric 
mass-squared differences ($\Delta{m}_{\odot}^{2}$ and $|\Delta{m}_{A}^{2}|$ 
respectively) defined by 
$r=\frac{\Delta{m}_{\odot}^{2}}{|\Delta{m}_{A}^{2}|}$ as exercised in
\cite{Hagedorn:2009jy,Altarelli:2009kr}. These mass-squared differences are
defined as $\Delta{m}_{\odot}^{2}=\Delta{m_{21}^{2}}={m_{2}^{2}-m_{1}^{2}}$ and
$|\Delta{m}_{A}^{2}|=|\Delta{m}^2_{31}|={m_{3}^{2}-m_{1}^{2}}
\approx|\Delta{m}^2_{32}|={m_{3}^{2}-m_{2}^{2}}$. Following 
\cite{Forero:2014bxa}, the best fit values of $\Delta
m^2_{\odot}=7.60\times10^{-5}\hspace{.1cm} \eVq$ (for both NH and IH) and
$|\Delta m^2_{A}|=2.48\times10^{-3}\hspace{.1cm}\eVq$ [NH] 
(and $|\Delta m^2_{A}|=2.38\times10^{-3}\hspace{.1cm}\eVq$ [IH]) will be used
in our analysis. Using Eqs.({\ref{M1real} - {\ref{M3real}} and {\ref{mi}}) we
obtain $r$ in terms of parameters involved in our framework as given by
\begin{equation}
 r=\frac{[\lambda^{2}_{2}+2\lambda_2{\rm K}\cos\phi_{ba}+{\rm 
K}^2-(1+\lambda_1)^2]
 (\lambda^{2}_{2}-2\lambda_2{\rm K}\cos\phi_{ba}+{\rm K}^{2})}
{4(1+\lambda_1)^2\lambda_2{\rm K}|\cos\phi_{ba}|}\label{rr}.
\end{equation}

\begin{figure}[h]
\begin{center}
\includegraphics[scale=0.7]{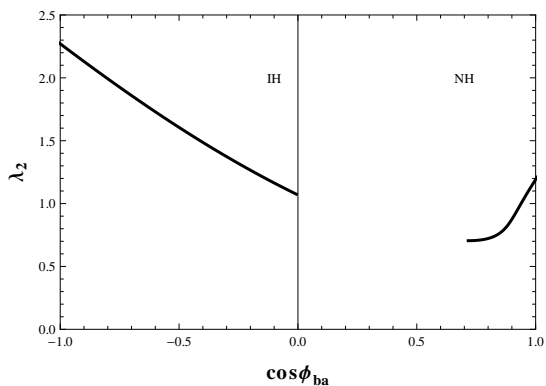}
\caption{{\small Variation of $\lambda_2$ with $\cos\phi_{ba}$. Here we have
fixed $\lambda_1$=0.37 for NH and $\lambda_1$=0.38 for IH.}}
\label{csa2}
\end{center}
\end{figure}

\noindent
We recall that $\phi_{ba}$ is the relative phase between parameters $b$ and
$a$. Note that with $\lambda_1=0$, K becomes unity and the expression for $r$
gets back the form in \cite{Altarelli:2009kr}. Considering $\lambda_1<1$ (as
required for $\theta_{13}$ being in the acceptable range, see Fig.{\ref{fig1}})
and as $\phi_{da}=0$, K becomes real and considered to be positive. Then as is
evident from Eq.({\ref{M1real}} - {\ref{M3real}}) and Eq.({\ref{mi}}),
$\cos\phi_{ba}>0$ for NH and $\cos\phi_{ba}<0$ for IH. Using $r=0.03$
\cite{Forero:2014bxa}, we can use Eq.({\ref{rr}}) now to study the 
correlation between $\lambda_2$ and $\cos\phi_{ba}$ as shown in 
Fig.{\ref{csa2}}. In doing so, we have set the value of $\lambda_1$ to be 0.37 
(0.38) which corresponds to the best fit value of $\sin^2\theta_{13}$ for NH 
(IH) as stated before. Obviously the right panel of the plot corresponds to NH
(as $\cos\phi_{ba}>0$) and left panel is for IH (as $\cos\phi_{ba}<0$). We find
that for NH, with $\lambda_1=0.37$, $\lambda_2$ is restricted to be in the
range $0.71-1.2$ and for IH, with  $\lambda_1=0.38$, $\lambda_2$ falls 
within\footnote[1]{Eq.(\ref{rr}) describes a quadratic
equation of $|\cos\phi_{ba}|$ once other parameters are fixed. The range of
$\lambda_2$ between 0 and 0.71 is excluded to keep the discriminant 
positive for $\lambda_1$=0.37 (for NH).} the range $1.1-2.3$. It will be
further modified as we proceed after including the constraint on the sum
of all the light neutrinos from the Planck data \cite{Ade:2013zuv}.

The light neutrino mass $m_{1}$ in this framework can be expressed as
\begin{equation}
m^{2}_{1}=|\Delta{m}_{A}^{2}|r\frac{(1+\lambda_1)^2}
{[\lambda^{2}_{2}+2\lambda_2{\rm K}\cos\phi_{ba}+{\rm 
K}^2-(1+\lambda_1)^2]}.
\label{m1}
\end{equation}
Now using the best fit value of $|\Delta{m}_{A}^{2}|=2.48\times{10^{-3}} {\rm
eV^2}$ [NH] ($2.38\times{10^{-3}} {\rm eV^2}$ [IH]), $r=0.03$ and
$\lambda_1=0.37$ (0.38), we can estimate $m_1$ from the above relation for NH
(IH), shown in Fig.{\ref{mia2}}, as a function of $\lambda_2$. 
\begin{figure}[h]
\centering
\includegraphics[scale=0.7]{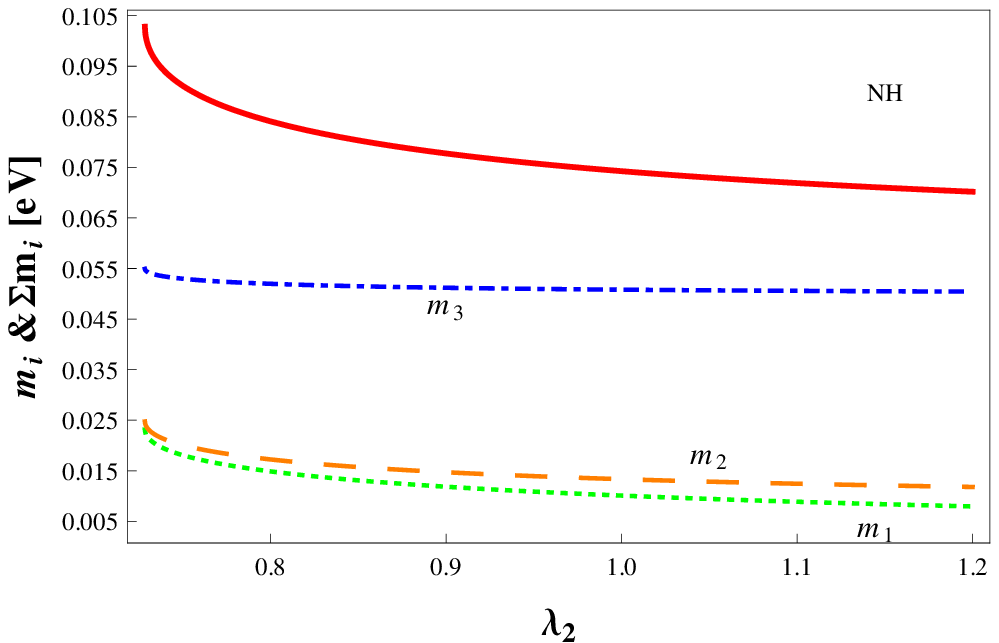}
\hspace{1cm}
\includegraphics[scale=0.7]{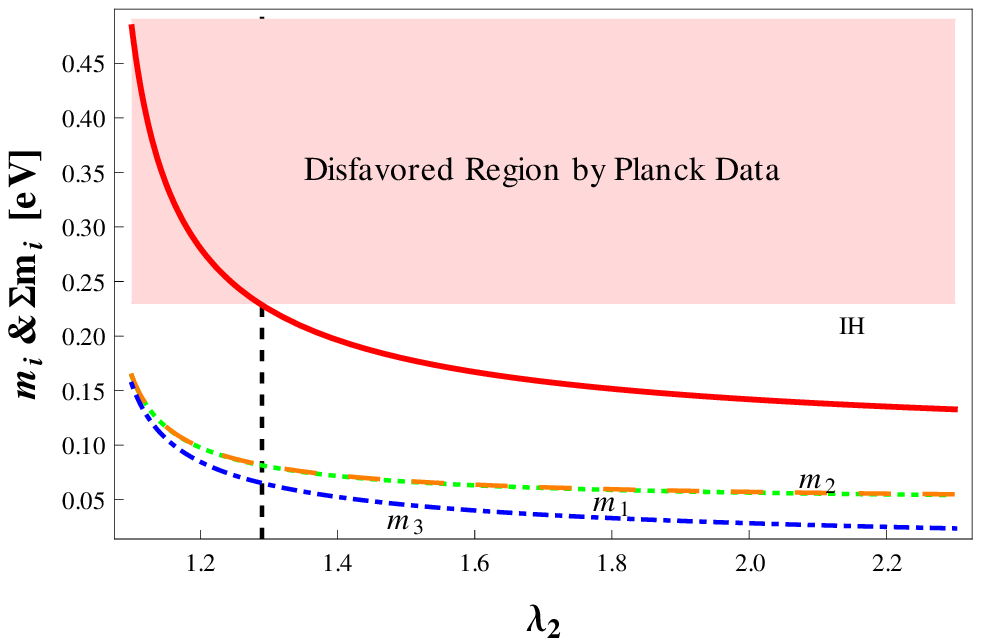}
\caption{{\small Light neutrino masses $m_{i}$'s and their sum, $\sum{m_{i}}$,
          as a function of $\lambda_2$ for NH ($\lambda_1=0.37$) and IH
          ($\lambda_1=0.38$). Here in the right panel the shaded region
          indicates the disfavored values of $\sum{m_{i}}$. This makes allowed
          range for $\lambda_2$ more restricted for IH, indicated by the
          vertical black dashed line.}}
\label{mia2}
\end{figure}
Similarly $m_{2}$ and $m_{3}$ are also plotted in Fig.{\ref{mia2}}. Note
that in doing this, the correct sign of $\cos\phi_{ba}$ in Eq.(\ref{m1}) needs 
to be taken into account while NH and IH cases are considered. The 
lightest neutrino mass $m_1$ ($m_3$) falls in the range 0.008 eV 
$\lesssim m_1\lesssim$ 
0.02 eV for NH (0.02 eV $\lesssim m_3 \lesssim$ 0.12 eV for IH). In this plot 
we 
have also shown the sum of the light neutrino masses, $\sum{m_i}$. From 
Fig.\ref{mia2} , we conclude that it lies in the range 0.07 eV 
$\lesssim\sum{m_i}\lesssim$ 0.1 eV for NH and 0.13 eV $\lesssim 
\sum{m_i}\lesssim$ 0.28 eV for IH. The Planck
data along with external CMB and BAO results \cite{Ade:2013zuv} provide an
upper bound as $\sum{m_i}\lesssim0.23$ eV. Once this is considered, the range of
$\sum{m_i}$ as obtained from our analysis for NH would not be affected. However
in case of IH, it further restricts the range of $\lambda_2$ 
($1.3\lesssim\lambda_2\lesssim2.3$, indicated by vertical dashed line) as 
observed from the shaded region of Fig.{\ref{mia2}},
right panel. So the model's prediction for sum of all three light neutrino
masses turns out to be, 
\begin{equation}
 0.07 \hspace{.1cm}{\rm eV} \lesssim \sum\limits_{i=1}^3 m_i \lesssim 0.1
 \hspace{.1cm}{\rm eV \hspace{.05cm}( NH)}  \hspace{.2cm} \&
\hspace{.2cm}
0.13 \hspace{.1cm}{\rm eV} \lesssim \sum\limits_{i=1}^3 m_i \lesssim 0.23
\hspace{.1cm}{\rm eV \hspace{.05cm}( IH).} 
\end{equation}
\noindent
In our analysis we can comment also on the relative magnitudes of heavy RH 
neutrinos. For NH we obtain $M_1\simeq(1.1-1.5)M_2\simeq(2.7-6.6)M_3$ and 
for IH we have $M_1\simeq{M_2}\simeq\frac{M_3}{1.2-2.3}$. So, in the present 
set-up Majorana neutrinos are not strongly hierarchical. 

Two Majorana phases $\alpha_{21}$ and $\alpha_{31}$ can be investigated in the 
set-up in a similar way as done in \cite{Hagedorn:2009jy}. Here in the model
under consideration, we find Majorana phases $\alpha_{21}$ and $\alpha_{31}$ in
terms of $\lambda_1$, $\lambda_2$ and $\phi_{ba}$ as given by
\begin{eqnarray}
\tan\alpha_{21}&=&-\frac{\lambda_2\sin\phi_{ba}}{{\rm
K}+\lambda_2\cos\phi_{ba}},\label{ta21}\\
\tan\alpha_{31}&=&\frac{2{\rm K}\lambda_2\sin\phi_{ba}}{\lambda^{2}_{2}-{\rm
K}^{2}}.\label{ta31}\\\nonumber
\end{eqnarray}

\begin{figure}[!htb]
\centering
{
      \includegraphics[scale=0.7]{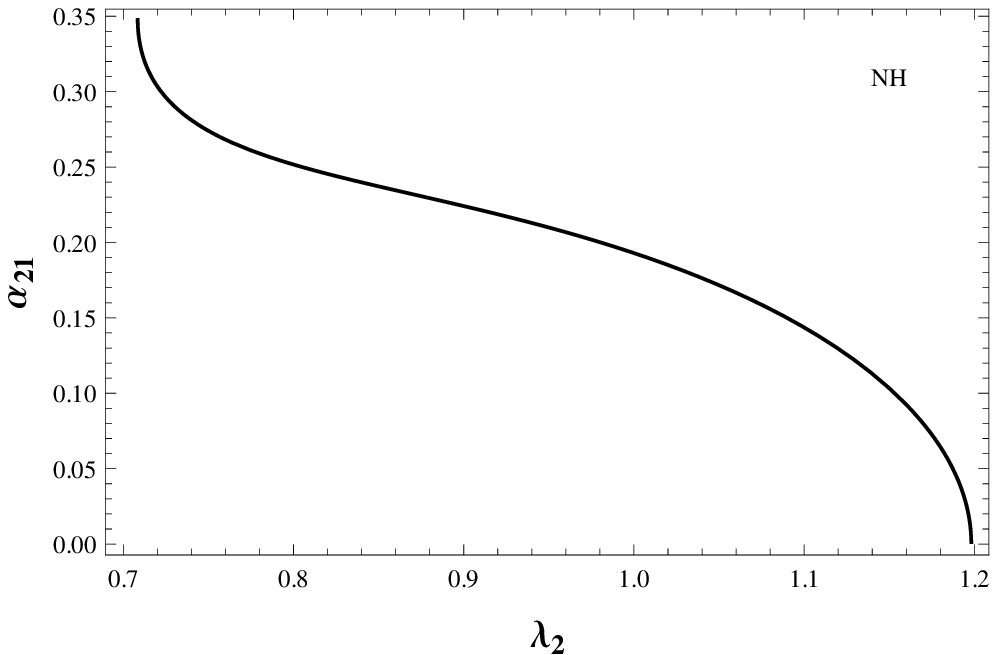}
      }
      \hspace{.5cm}
    {
     \includegraphics[scale=0.7]{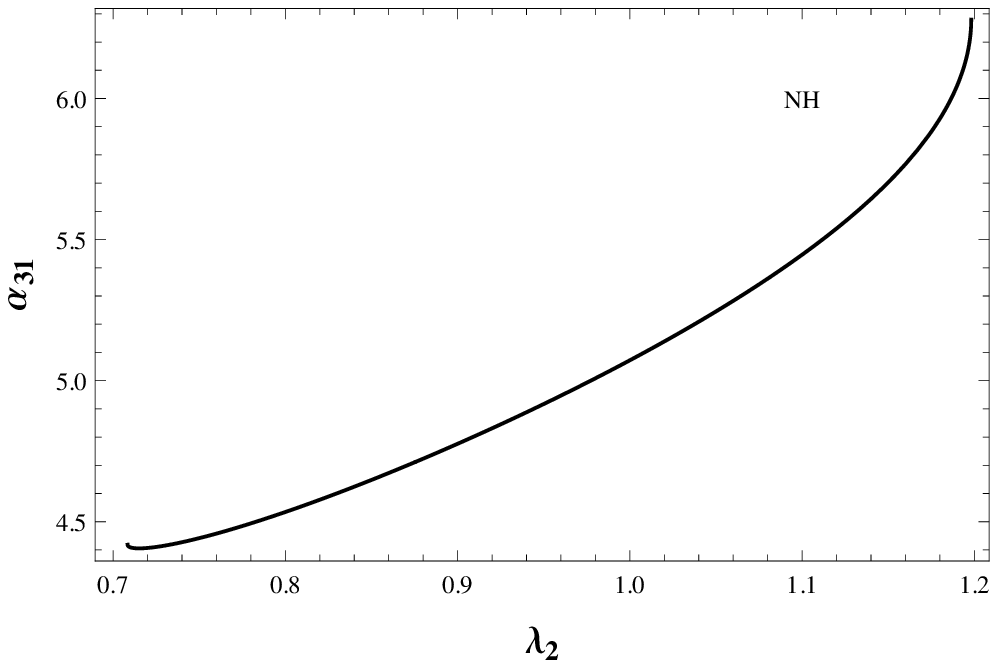}
      }
\caption{{\small  Variation of Majorana phases ($\alpha_{21}$: left panel; 
                  $\alpha_{31}$: right panel) with $\lambda_2$ for NH.}}
\label{mpnh}
\end{figure}
\noindent
Note that there exists a relative sign between $\sin\alpha_{21}$ and 
$\sin\alpha_{31}$ as observed from the neutrino mass sum rule in
Eq.({\ref{sumrule}}). For NH, $\cos\phi_{ba}>0$ as discussed before and 
$\sin\phi_{ba}<0$ is considered in order to produce correct sign of baryon
asymmetry \cite{Hagedorn:2009jy}. Similarly, for IH we have $\cos\phi_{ba}<0$
and $\sin\phi_{ba}<0$. Taking all this into consideration, Eqs.(\ref{ta21}) and 
(\ref{ta31}) can successfully correlate Majorana phases ($\alpha_{21}$ and 
$\alpha_{31}$) with parameters $\lambda_1$ and $\lambda_2$. We have plotted
variation of $\alpha_{21}$ and $\alpha_{31}$ with $\lambda_2$ for both NH and IH
in Fig.\ref{mpnh} and Fig.\ref{mpih} respectively. As before we have fixed
$\lambda_1=0.37$ for NH ($\lambda_1=0.38$ for IH).
\begin{figure}[!htb]
\centering
{
      \includegraphics[scale=0.7]{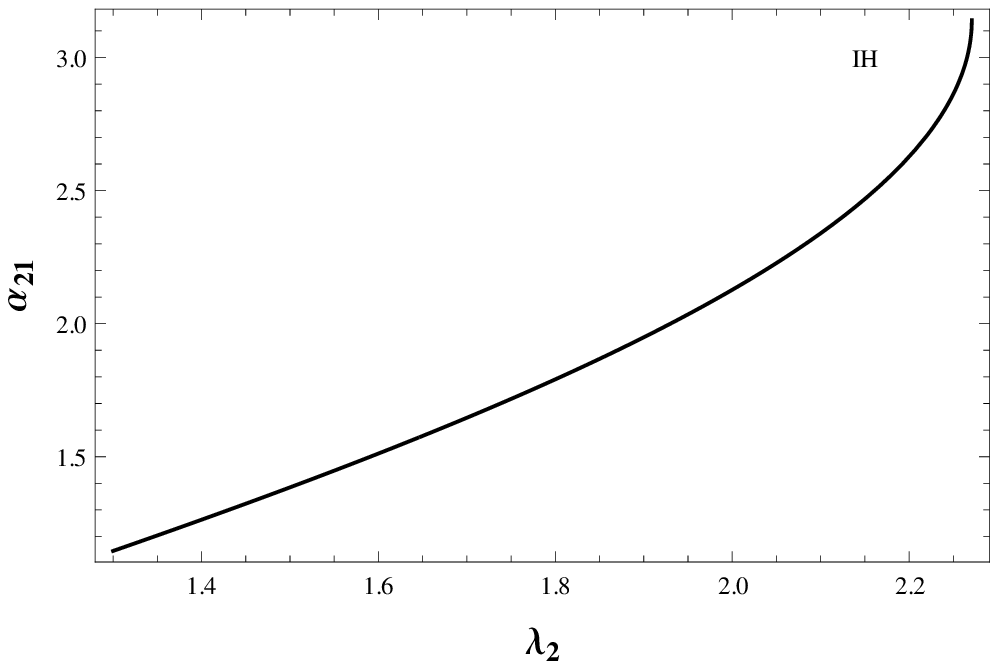}
      }
      \hspace{.5cm}
     {
      \includegraphics[scale=0.7]{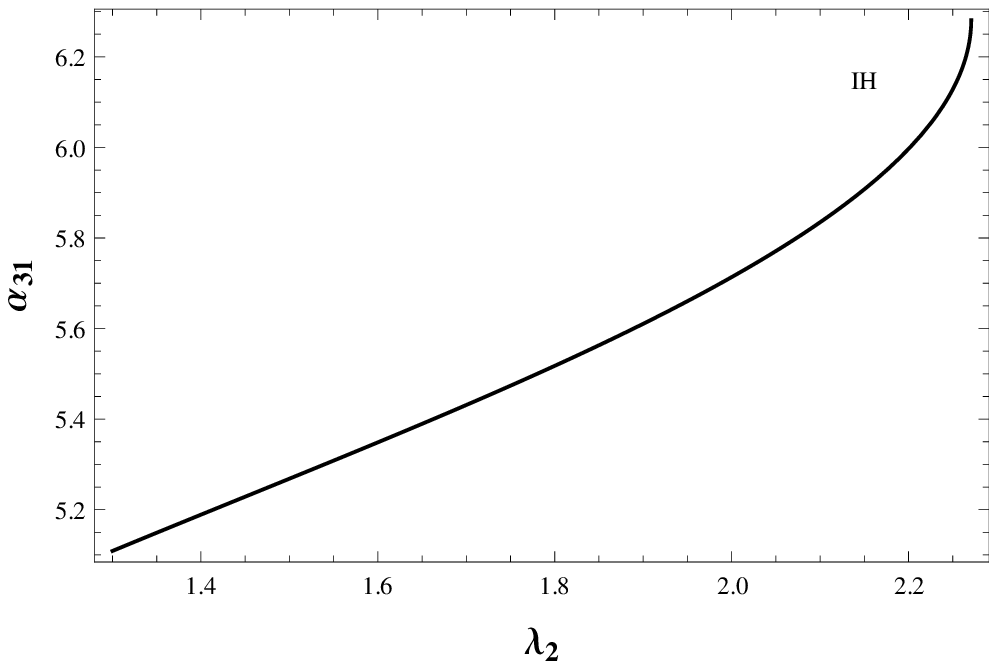}
      }
\caption{{\small Variation of Majorana phases ($\alpha_{21}$: left panel; 
                  $\alpha_{31}$: right panel) with $\lambda_2$ for IH.}}
\label{mpih}
\end{figure}
This study of Majorana phases will be particularly useful when we will study
the dependence of CP-violating parameter $\epsilon_i$ in our model on
$\lambda_2$. Effective neutrino mass parameter, $|\langle{m}\rangle|$, is an
important quantity which controls the neutrinoless double beta decay. In our 
model, the effective neutrino mass parameter is obtained
as \cite{Beringer:1900zz,Bilenky:2012qi}
\begin{equation}
 \left|\langle{m}\rangle\right|=\left|\frac{2}{3}m_{1}\cos^2\theta+\frac{1}{3}m_
{2}e^{i\alpha_{21}}+\frac{2}{3}m_{3}\sin^{2}\theta{e^{i\alpha_{31}}}\right|.
 \label{meenh}
\end{equation}
Since the dependence of $m_i$ and $\alpha_{21,31}$ on $\lambda_2$ (for fixed
$\lambda_1$) is known (from Fig.\ref{mia2}, \ref{mpnh} and \ref{mpih}), we plot
$|\langle{m}\rangle|$ as a function of $\lambda_2$ with ${\lambda_1}=0.37$ for 
NH and ${\lambda_1}=0.38$ for IH in Fig.{\ref{meenhp1}}. 
\begin{figure}[h]
\begin{center}
\includegraphics[scale=0.7]{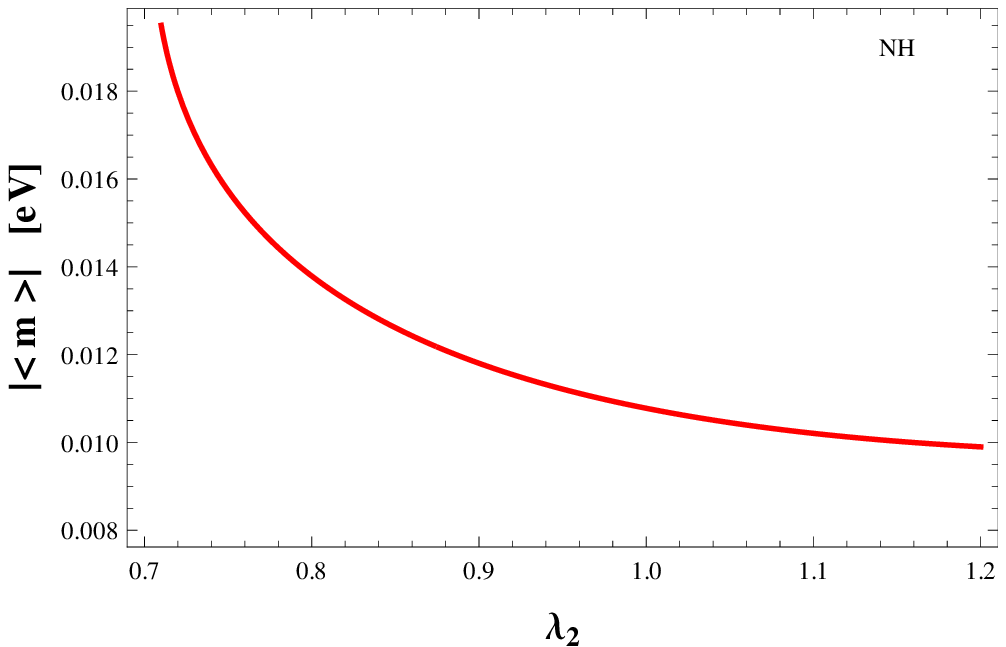}
\hspace{.5cm}
\includegraphics[scale=0.7]{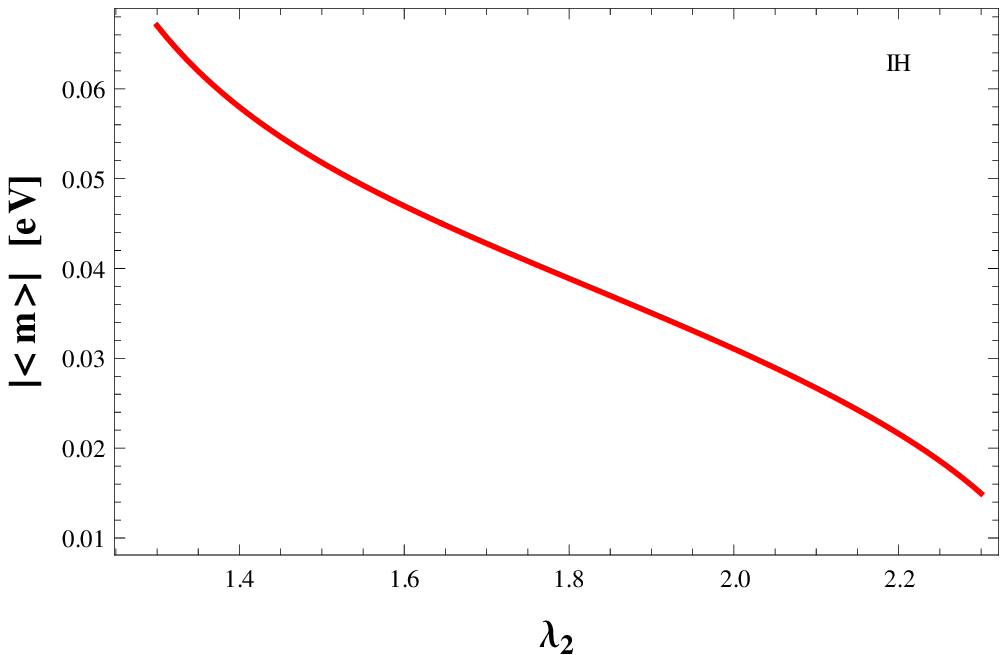}
\caption{{\small Variation of $\left|\langle{m}\rangle\right|$  with $\lambda_2$
        for NH (left panel) and IH (right panel) respectively.}}
\label{meenhp1}
\end{center}
\end{figure}
We found the range for the $\left|\langle{m}\rangle\right|$ as
$0.01\hspace{.1cm}{\rm
{eV}}<\left|\langle{m}\rangle\right|<0.02\hspace{.1cm}{\rm {eV}}$ for NH and 
$0.015\hspace{.1cm}{\rm{eV}}<\left|\langle{m}\rangle\right|<
0.067\hspace{.1cm}{\rm {eV}}$ for IH. The current upper limit on
$\left|\langle{m}\rangle\right|$ however varies between $0.177$ eV and $0.339$
eV taking into account the different choices of nuclear matrix elements
\cite{Huang:2014qwa}.
%%%%%%%%%%%%%%%%%%%%%%%%%%%%%%%%%%%%%%%%%%%%%%%
%%%%%%%%%%%%%%				 %%%%%%%%%%%%%%%%%%%%%%%
%%%%%%%%%%%%%	         Section 5        %%%%%%%%%%%%%%%%%%%%%%%
%%%%%%%%%%%%%%				  %%%%%%%%%%%%%%%%%%%%%%
%%%%%%%%%%%%%%%%%%%%%%%%%%%%%%%%%%%%%%%%%%%%%%%
\section{Leptogenesis}\label{lep}
The presence of see-saw realization of light neutrino mass in the model 
under consideration gives the opportunity to study leptogenesis as the heavy RH
neutrinos are already present in the model. It allows the 
generation of lepton asymmetry through the out-of-equilibrium decay of heavy RH 
neutrinos in the early Universe. This lepton asymmetry can be converted into 
baryon asymmetry of the Universe with the help of sphaleron process. With the 
consideration that the generation of lepton asymmetry happens at a 
temperature of the Universe $T\sim{M_i}\gtrsim(1+\tan^2\beta)10^{12}$ GeV
(where $\tan\beta=v_u/v_d$), it does not distinguish between flavors, the so
called `one-flavor approximation' regime 
\cite{Hagedorn:2009jy,Pascoli:2006ie, Blanchet:2006ch,Davidson:2008pf}
is achieved. The CP-asymmetry generated by the out-of-equilibrium decay of each
RH neutrinos (and sneutrinos) is given by
\cite{Fukugita:1986hr, Flanz:1994yx, Covi:1996wh, Plumacher:1996kc, 
Buchmuller:1997yu, Buchmuller:2004nz,oai:arXiv.org:hep-ph/0310123} 
\begin{equation}\label{epsi}
 \epsilon_{i}=\frac{1}{8\pi}\sum_{j\ne{i}}
 \frac{{\rm 
Im}\left[\left((\hat{Y}_{\nu}\hat{Y}^{\dagger}_{\nu})_{ji}\right)^{2}\right]}
 {(\hat{Y}_{\nu}\hat{Y}^{\dagger}_{\nu})_{ii}}f\left(\frac{m_{i}}{m_{j}}\right),
\end{equation}
where $\hat{Y}_{\nu}$ is the effective Yukawa coupling matrix for neutrinos in 
the basis where RH neutrino mass matrix $M_{Rd}$ is diagonal
\footnote[2]{Here Eq.(\ref{mi}) is used to express the loop factor 
$f$ in terms of the ratio of light neutrino masses.}. In the present 
set-up, 
$\hat{Y}_{\nu}={\rm
diag}(1,e^{-i\alpha_{21}/2},e^{-i\alpha_{31}/2})U_R^TY_{\nu}$ , where 
$U_R=U_{TB}U_{1}$. The loop factor $f(x)$ in the above expression (the model 
being supersymmetric) is defined as follows 
\cite{oai:arXiv.org:hep-ph/0310123}
\begin{equation}
 f(x)\equiv{-x}\left[\frac{2}{x^{2}-1}+\ln\left(1+\frac{1}{x^{2}}\right)\right]
,
\end{equation}
with $x=m_{i}/m_{j}$. The total lepton asymmetry receives contribution from the 
decay of all three RH neutrinos (and sneutrinos). 

It has been observed that at LO, ({\textit{i.e.} when $Y_{\nu}=Y_{\nu0}$)
product of the effective Yukawa coupling matrices 
$\hat{Y}_{\nu{0}}\hat{Y}^{\dagger}_{\nu{0}}$ is proportional to a unit matrix, 
hence lepton asymmetry parameter $\epsilon_{i}$ vanishes \cite{Jenkins:2008rb}. 
However considering NLO corrections to the Yukawa, we have obtained
Eq.(\ref{dynu}). Therefore using Eq.(\ref{dynu}), 
$\hat{Y}_{\nu}\hat{Y}^{\dagger}_{\nu}$ becomes 
{\small 
\begin{eqnarray}
\hat{Y}_{\nu}\hat{Y}^{\dagger}_{\nu}=y^{2}{\bf I}+&
\begin{bmatrix}\label{ynuynud}
              \cos{2\theta} & \sqrt{2}e^{\frac{i\alpha_{21}}{2}}\cos{\theta} 
& 
e^{\frac{i\alpha_{31}}{2}}\sin{2\theta}\\
              \sqrt{2}e^{-\frac{i\alpha_{21}}{2}}\cos{\theta} & 0 & 
\sqrt{2}e^{\frac{i(\alpha_{31}-\alpha_{21})}{2}}\sin{\theta}\\
              e^{\frac{-i\alpha_{31}}{2}}\sin{2\theta} & 
\sqrt{2}e^{\frac{-i(\alpha_{31}-\alpha_{21})}{2}}\sin{\theta} & -\cos{2\theta}
              \end{bmatrix}\left(2{\rm Re}(x_{C})\kappa{y}\right)\\
             & +\begin{bmatrix}
              -\frac{\sin{2\theta}}{\sqrt{3}} & 
\sqrt{\frac{2}{3}}e^{i\frac{i\alpha_{21}}{2}}\sin{\theta}  & 
\frac{1}{\sqrt{3}}e^{\frac{i\alpha_{31}}{2}}\cos{2\theta}\\
              \sqrt{\frac{2}{3}}e^{-i\frac{i\alpha_{21}}{2}}\sin{\theta} & 0 & 
-\sqrt{\frac{2}{3}}e^{\frac{i(\alpha_{31}-\alpha_{21})}{2}}\cos{\theta}\\
              \frac{1}{\sqrt{3}}e^{\frac{-i\alpha_{31}}{2}}\cos{2\theta} & 
-\sqrt{\frac{2}{3}}e^{\frac{-i(\alpha_{31}-\alpha_{21})}{2}}\cos{\theta} & 
\frac{1}{\sqrt{3}}\sin{2\theta}
              \end{bmatrix}\left(2{\rm Re}(x_{D})\kappa{y}\right).\nonumber
\end{eqnarray}
}Note that having origin related to a NLO correction term, $\kappa$ in 
general is expected to be small, $\kappa=v_T/\Lambda\ll1$. Hence the expression 
of Eq. (\ref{ynuynud}) is kept up to first order in $\kappa$. Finally in our 
framework the CP-asymmetry parameters corresponding to all three RH neutrinos, 
$\epsilon_{1,2,3}$ take the form as 
{\small
\begin{eqnarray}
\epsilon_{1} = 
\frac{-\kappa^2}{2\pi}\left[\sin\alpha_{21}
\left(2{\rm Re} (x_{C})^{2}
 \cos^{2}{\theta} + \frac{2{\rm Re}(x_{D})^{2}}{3} \sin^{2}{\theta} + 
\frac{2{\rm Re}(x_{C}){\rm Re}(x_{D})}{\sqrt{3}} \sin{2\theta}\right)
 f\left(\frac{m_{1}}{m_{2}}\right)\right.\nonumber\\
 \left.+\sin\alpha_{31}\left({\rm Re}(x_{C})^{2}\sin^{2}{2\theta} + \frac{{\rm 
Re}(x_{D})^{2}}{3} \cos^{2}{2\theta} +
  \frac{{\rm Re}(x_{C}){\rm Re}(x_{D})}{\sqrt{3}} 
\sin{4\theta}\right)f\left(\frac{m_{1}}{m_{3}}\right)\right],\label{epi0}
\end{eqnarray}

\begin{eqnarray}
\epsilon_{2}=\frac{\kappa^2}{2\pi}\left[
\sin\alpha_{21}\left(2{\rm Re}(x_{C})^{2}
 \cos^{2}{\theta}+\frac{2{\rm Re}(x_{D})^{2}}{3} \sin^{2}{\theta}+\frac{2{\rm 
Re}(x_{C}){\rm Re}(x_{D})}{\sqrt{3}}\sin{2\theta}\right)
 f\left(\frac{m_{2}}{m_{1}}\right)\right.\nonumber\\
 \left.-\sin(\alpha_{31}-\alpha_{21})\left(2{\rm 
Re}(x_{C})^{2}\sin^{2}{\theta}+\frac{2{\rm Re}(x_{D})^{2}}{3} \cos^{2}{\theta} -
  \frac{2{\rm Re}(x_{C}){\rm Re}(x_{D})}{\sqrt{3}} \sin{2\theta} 
\right)f\left(\frac{m_{2}}{m_{3}}\right)\right],\label{epi1}
\end{eqnarray}

\begin{eqnarray}
\epsilon_{3} = 
\frac{\kappa^2}{2\pi}\left[\sin\alpha_{31}
\left({\rm Re}(x_{C})^{2}
 \sin^{2}{2\theta} + \frac{{\rm Re}(x_{D})^{2}}{3} \cos^{2}{2\theta} + 
\frac{{\rm Re}(x_{C}){\rm Re}(x_{D})}{\sqrt{3}} \sin{4\theta}\right)
 f\left(\frac{m_{3}}{m_{1}}\right)\right.\nonumber\\
\left. +\sin(\alpha_{31}-\alpha_{21}) \left(2{\rm
Re}(x_{C})^{2}\sin^{2}{\theta} 
+\frac{2{\rm Re}(x_{D})^{2}}{3} \cos^{2}{\theta} -
  \frac{2{\rm Re}(x_{C}){\rm Re}(x_{D})}{\sqrt{3}} \sin{2\theta} 
\right)f\left(\frac{m_{3}}{m_{2}}\right)\right]\label{epi2}.
\end{eqnarray}
}Lepton asymmetry in this scenario therefore depends on light neutrino masses 
$m_i$ (through loop factor), Majorana phases $\alpha_{21,31}$, couplings 
 Re($x_{C,D}$), $\kappa$ (coming from the NLO correction terms in Yukawa) 
and interestingly on $\theta$ (and hence on $\lambda_1$). Recall that $\theta$ 
was originated from the deviation from the exact tri-bimaximal mixing and 
therefore leads to nonzero $\sin{\theta_{13}}$. We will come back to discuss it,
 before that let us discuss how this lepton asymmetry 
parameter is connected with observed baryon asymmetry of the Universe.

Lepton asymmetry can be linked to the baryon asymmetry \cite{Davidson:2008bu, 
Harvey:1990qw, Engelhard:2007kf,Engelhard:2006yg} as 
\begin{equation}
{Y_{B}}={-1.48\times{10^{-3}}}\sum_i\epsilon_{i}\eta_{ii}.
\end{equation}
Here $\eta_{ii}$ stands for the efficiency factor 
\cite{oai:arXiv.org:hep-ph/0310123}. We consider the efficiency factor to be 
given by
\begin{equation}\label{eta}
 \frac{1}{\eta_{ii}} \approx \frac{3.3 \times 10^{-3}\hspace{.1cm} {\rm eV} 
}{\tilde{m_i}}+\left(\frac{\tilde{m_i}}{0.55 \times
10^{-3}\hspace{.1cm} {\rm eV}} \right)^{1.16},
\end{equation}
\noindent
with $\tilde{m_{i}}$ as the washout mass parameter, $\tilde{m_{i}}= 
\frac{(\hat{Y}_{\nu}\hat{Y}^{\dagger}_{\nu})_{ii}v_u^2}{M_{i}}\simeq{m_i}$ to 
the leading order. The above expression is valid for $M_i<10^{14}$ GeV
\cite{Petcov:2003zb}. This upper bound on $M_i$ is also consistent in keeping
the lepton number violating decays  within the experimental 
limit\cite{Petcov:2003zb, Petcov:2005jh}. As we already have a lower bound on 
$M_{i}$ from the `one-flavor approximation', it turns out that low values of 
$\tan\beta$ are favored for this scenario to work\footnote[3]{$y$ is expected 
to be $\sim\mathcal{O}(10^{-1})$ in order to reproduce correct $m_i$ for this 
range of $M_{i}$.}. Interestingly in \cite{Djouadi:2013vqa}, the authors have 
shown that if the scale of supersymmetry breaking ($m_s$) in MSSM is 
sufficiently large (say $\sim$ 10 TeV or so) the low $\tan\beta$ region
$\tan\beta\lesssim(3-5)$ is consistent with the results of LHC so far. Such 
large value of $m_s$ on the other hand can in principle reduce the branching 
ratio for the LFV processes. However the details of this conjecture is beyond 
the scope of the present study.

\subsection{Leptogenesis with fixed $\lambda_1$ and varying $\lambda_2$}
In this section we will study the range of the parameters involved in $Y_B$ 
expression so as to reproduce the correct amount of matter-antimatter asymmetry 
of the Universe. The observed value of $Y_B$ is reported to be
\cite{Bennett:2012zja}
\begin{equation}
 Y_{B}=(8.79 \pm 0.20)\times 10^{-11}.
\end{equation}
As the efficiency factor ($\eta_{ii}$) is found to be
$\sim{\mathcal{O}(10^{-2})}$, $\epsilon_i$ should be of order
$\mathcal{O}(10^{-6})$ in order to reproduce the correct amount of baryon
asymmetry of the Universe. As discussed earlier, we have kept $\lambda_1$ to be
fixed at 0.37 for NH (0.38 for IH) which correspond to the best fit value of 
$\sin^2\theta_{13}$ \cite{Forero:2014bxa}. We further note that the expression
of $Y_{B}$ involves $\theta$ which in tern is related to $\theta_{13}$. So
once $\lambda_1$ is fixed it would correspond to a particular value of $\theta$.
The expansion parameter $\kappa=v_T/\Lambda$ is taken to be $\sim10^{-2}$. The
variation of $\alpha_{21,31}$ and $m_i$'s with $\lambda_2$ (for
$\lambda_1=0.37$, $0.38$ for NH and IH respectively) are already studied. Using
those information, we can study the dependence of $Y_{B}$ on $\lambda_2$ for
fixed values of Re$(x_{C})$ and Re$(x_{D})$. The first 
bracketed expression in Eqs.(\ref{epi0} - \ref{epi2}) therefore serves merely 
as constant factor.

\begin{figure}[h]
\centering
\includegraphics[scale=0.7]{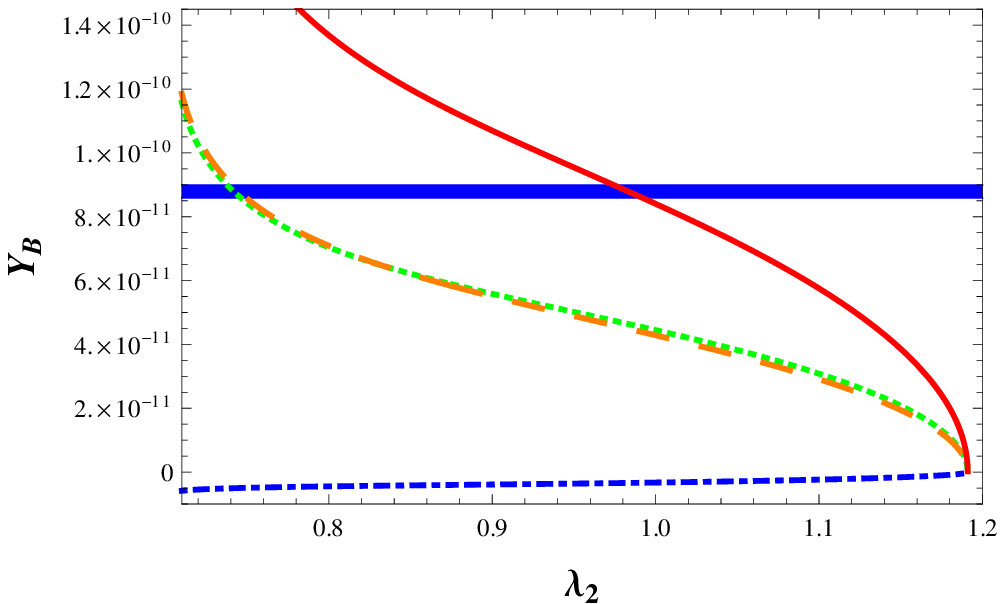}
\hspace{1cm}
\includegraphics[scale=0.7]{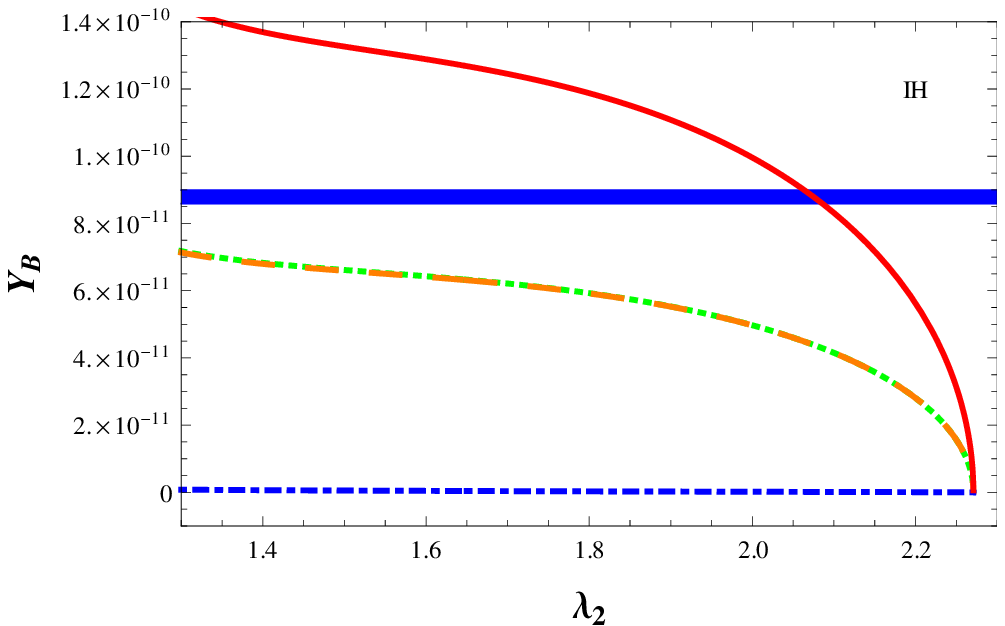}
\caption{{\small Baryon asymmetry of the Universe as function of
                 $\lambda_2$ for NH (with $\lambda_1=0.37$, left panel) and IH
                 (with $\lambda_1=0.38$, right panel). Here, red continuous
                 line, orange large dashed line, green dotted line and blue
                 dot-dashed line stand for total $Y_{B}$, $Y_{B1,2,3}$
                 respectively. The horizontal blue patch represents
                 allowed range for total baryon asymmetry. For NH we have taken
                 Re$(x_C)$ = Re$(x_D)$=0.2 and for IH we have Re$(x_C)$ = 
                 Re$(x_D)$=0.05. For both cases we have fixed $\kappa$ at
                 0.01.}}
\label{figyba2}
\end{figure}

In Fig.\ref{figyba2} (left panel), we have plotted total baryon asymmetry
$Y_{B}$ (red continuous line) along with individual $Y_{B1,2,3}$ (orange large
dashed, green dotted and blue dot-dashed lines respectively) against $\lambda_2$
for Re$(x_{C})$=Re$(x_{D})=0.2$ in case of NH. Note that the range of
$\lambda_2$ $0.71-1.2$ for NH and $1.3-2.3$ for IH was already fixed (from Fig.
\ref{csa2} and \ref{mia2}) for $\lambda_1=0.37$ (for NH) and $\lambda_1=0.38$
(for IH) respectively. The relative sign between $\sin\alpha_{21}$ and
$\sin\alpha_{31}$ is fixed from the sum rule, Eq.(\ref{sumrule}). Their
dependence on $\lambda_2$ is depicted in Fig.\ref{mpnh}. In producing these
plots, we recall that $\cos\phi_{ba}>0$ for NH and $\cos\phi_{ba}<0$ for IH.
Also $\sin\phi_{ba}<0$ is considered to produce correct sign of $Y_{B}$. In
$\epsilon_1$, $f(m_1/m_2)$ is of positive sign and remains dominant over
$|f(m_1/m_3)|$ throughout the range of $\lambda_2$ by orders of magnitude. 
So an overall negative sign for $\epsilon_1$ results when combined with 
$\sin\alpha_{21}>0$ and $\sin\alpha_{31}<0$ for the range of $\lambda_2$ 
inferred from Fig.\ref{figyba2}. Similar conclusion can be 
drawn for $\epsilon_2$. In this case $f(m_2/m_1)$ is negative and its magnitude
is sufficiently large compared to $|f(m_2/m_3)|$ so that differences between
magnitude of $\sin\alpha_{21}$ and $\sin(\alpha_{31}-\alpha_{21})$ can not
produce any sizable effect between the two terms (one is the set of terms
proportional to $\sin\alpha_{21}$ and other is the similar set proportional to
$\sin(\alpha_{31}-\alpha_{21})$) involved. So $\epsilon_2$ is effectively
dominated by the first term and overall it gives negative contribution. In
$\epsilon_3$, however both the terms involved contribute almost equally and
overall $\epsilon_3$ contributes with opposite sign (also seen in the
Fig.\ref{figyba2}\hspace{.1cm} terms of $Y_{B3}$ which is negative, left panel) 
compared to $\epsilon_{1,2}$. As shown in Fig.\ref{figyba2} (left panel), 
contribution from $Y_{B3}$ is 
suppressed (and of opposite sign). This is due to the fact that the 
corresponding washout is larger though in magnitude 
$|\epsilon_3|\lesssim|\epsilon_{1,2}|$. A horizontal patch in Fig.\ref{figyba2}
is provided to indicate the allowed $Y_{B}$ range \cite{Bennett:2012zja}.
It shows that for this specific choice of Re$(x_{C,D})$=0.2, correct amount of 
baryon asymmetry can be generated in our framework for 
$\lambda_2\sim\mathcal{O}(1)$. Note that we can achieve this $Y_{B}$ for not so 
large value of Re$(x_{C,D})$ in comparison to the findings of
\cite{Hagedorn:2009jy}. To check the possible values of Re$(x_{C})$ and/or 
Re$(x_{D})$ we have drawn a contour plot in Fig.\ref{figxd} (left panel)
between Re$(x_{D})$ and $\lambda_2$, while Re$(x_{C})$=Re$(x_{D})$ is assumed 
as an example. The pattern of $Y_{B}$ plot is also different from what was 
obtained in \cite{Hagedorn:2009jy}. This is due to the involvement of nonzero 
$\theta$.

In Fig.\ref{figyba2} (right panel), we then plot $Y_{B}$, $Y_{B1,2,3}$ vs.
$\lambda_2$ in case of IH with Re$(x_C)$ = Re$(x_D)$ = 0.05. As it was found in
section \ref{sec4}, $\lambda_2$ ranges between 1.3 and 2.3 and
$\cos\phi_{ba}<0$ and $\sin\phi_{ba}<0$ in this case. The Majorana 
CP-violating phases $\alpha_{21}$ and $\alpha_{31}$ are obtained in section 
\ref{sec4} as function of $\lambda_2$ (see Fig.\ref{mpih}, with
$\lambda_1=0.38$). Here $m_1$ and $m_2$ are much closer to each other leading 
to large enhancement in the magnitude of loop factors $f(m_1/m_2)$ and 
$f(m_2/m_1)$. Their magnitudes are even larger than their counterpart in NH. 
Variation of these loop factors with $\lambda_2$ shows
that $f(m_1/m_2)\simeq{-}f(m_2/m_1)\gg{f(m_3/m_{1,2})}$ and 
$f(m_1/m_2)\simeq{-}f(m_2/m_1)\gg{-f(m_{2,1}/m_{3})}$. Overall nonzero
CP-violating phases $\alpha_{21}$ and $\alpha_{31}$ are required to have
leptogenesis but it appears that the final asymmetry is dominated by the 
loop factors. Though $Y_{B1}$ and $Y_{B2}$ face relatively large washout
effect, still they generate sizable contribution and $Y_{B3}$ gives sub-dominant
contribution as shown in Fig.{\ref{figyba2}}. Here also we have plotted a
contour between Re$(x_{D})$ and $\lambda_2$, assuming Re$(x_{C})$ = 
Re$(x_{D})$ with $Y_{B}$ fixed at its central value, as shown in Fig.\ref{figxd} 
(right panel). We find that in this case, smaller values of 
Re$(x_{C})$ = Re$(x_{D})$ are favored compared to the ones in NH case.

\begin{figure}[h]
\centering
\includegraphics[scale=0.6]{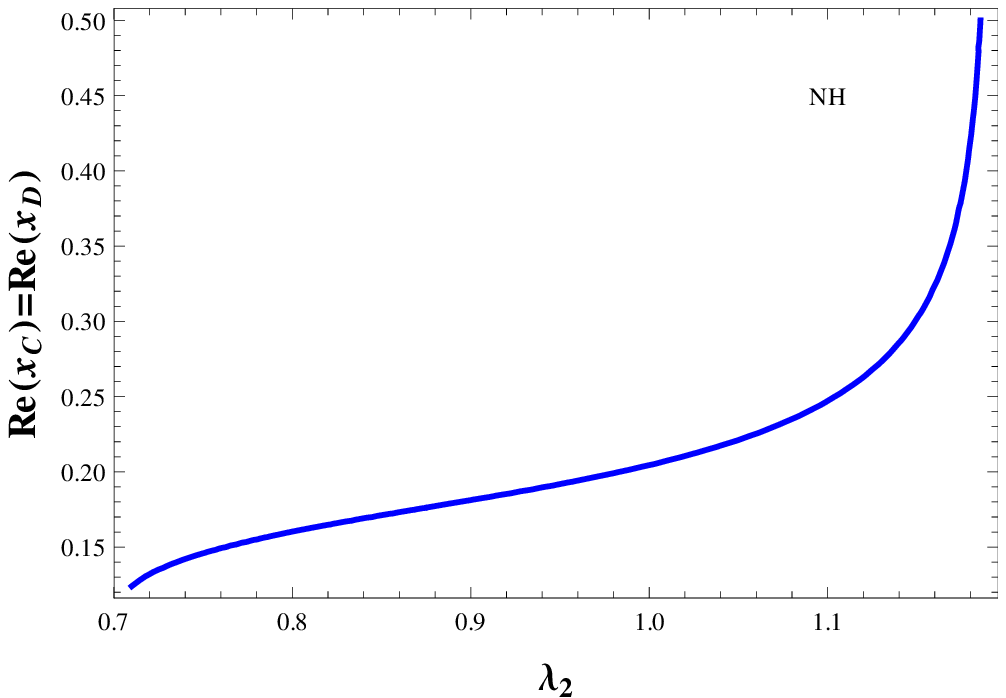}
\hspace{1cm}
\includegraphics[scale=0.6]{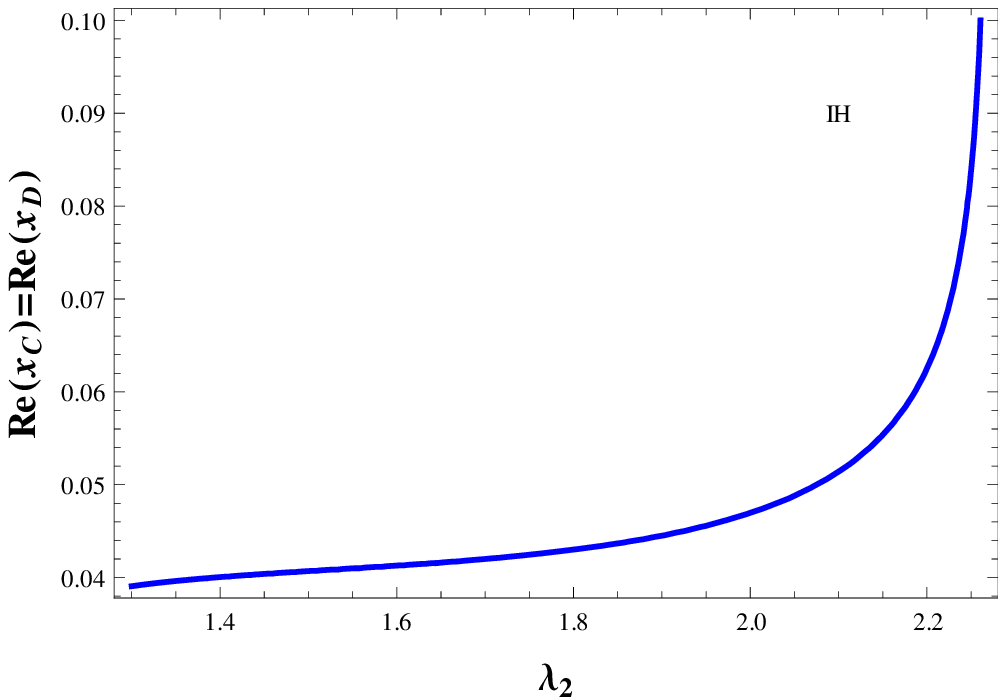}
\caption{{\small  Contour plot of Re$(x_{C})$(=Re$(x_{D})$) and $\lambda_2$,
                  with $Y_{B}$ fixed at its central value.}}
\label{figxd}
\end{figure}
Since the RH Majorana neutrino masses (in IH case particularly) are close to 
each other, we need to check the possibility of satisfying condition for 
resonant leptogenesis \cite{reso_lepto}. In our model, the quantity related to 
the mass degeneracy has been computed and found to be
\begin{equation}
 \frac{M_2}{M_1}-1\approx(10^{-2}-10^{-3}),
\end{equation}
\noindent
after scanning over the full range of $\lambda_2$
($1.3\lesssim\lambda_2\lesssim2.3$). We find that the resonance condition,
\begin{equation}
 \left|\frac{M_2}{M_1}-1\right|\sim 
\left|\frac{\left(\hat{Y}_{\nu}\hat{Y}^{\dagger}_{\nu}\right)_{12}}{16\pi}
\right|, \nonumber
\end{equation}
is not satisfied in our model. This is because the term in the right-hand-side 
of the resonance condition turns out to be of order 
$5\times10^{-2}\kappa{y}[{\rm {Re}}(x_C)\cos\theta+{\rm {Re}}(x_D)\sin\theta]$. 
As $\kappa\sim10^{-2}$, $y\sim10^{-1}$ and $\theta$ is expected to produce 
correct $\theta_{13}$,
\begin{equation} 
\frac{\left(\hat{Y}_{\nu}\hat{Y}^{\dagger}_{\nu}\right)_{12}}{16\pi}
\sim 10^{-5}-10^{-6}.\nonumber
\end{equation}
Hence, in the present model, the resonant condition is not satisfied. 
\subsection{Leptogenesis with fixed $\lambda_2$ and varying $\lambda_1$}
In this case we have taken a different approach by keeping $\lambda_2$ fixed 
at certain value, $\lambda_2=1$ for NH and $\lambda_2=2.1$ for IH
\footnote[4]{From Fig.\ref{csa2} and Planck limit on $\sum{m_i}$, note that 
there is no such common value of $\lambda_2$ exists for which both NH and IH 
cases can be considered.}. Then we can study the variation of $Y_{B}$ with
$\lambda_1$. The range of $\lambda_1$ (0.33$\lesssim\lambda_1\lesssim$0.41) is
of course restricted from Fig.\ref{fig1} in section \ref{sec3}. By using
Eq.(\ref{rr}) and taking $r=0.03$, we can now investigate the variation of
$\cos\phi_{ba}$ vs. $\lambda_1$. 
\begin{figure}[h]
\centering
\includegraphics[scale=0.6]{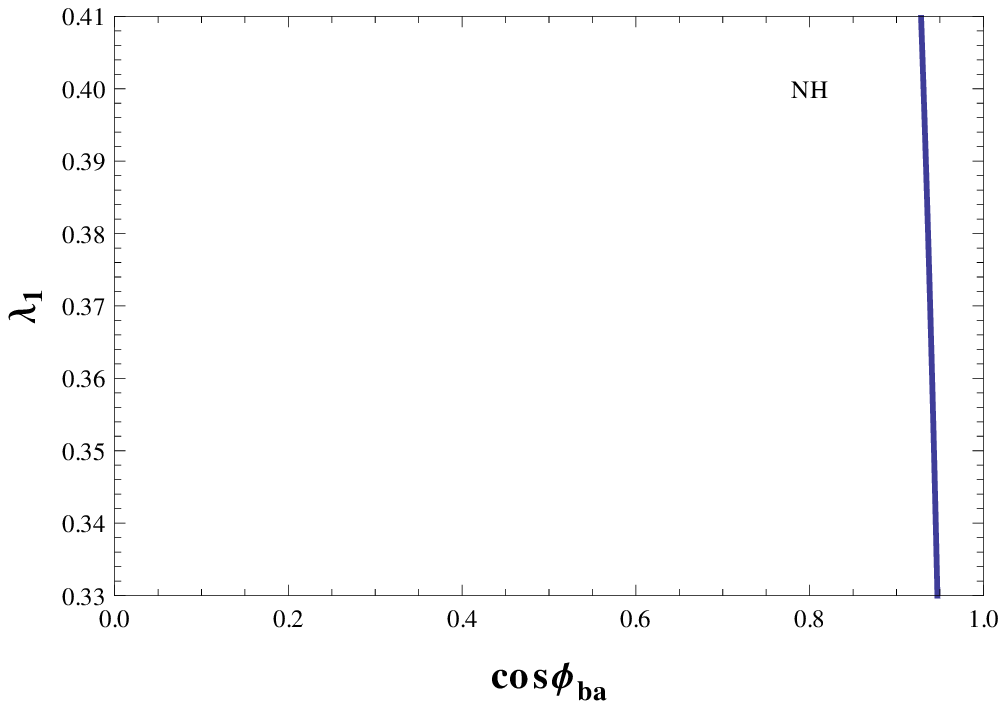}
\hspace{1.7cm}
\includegraphics[scale=0.6]{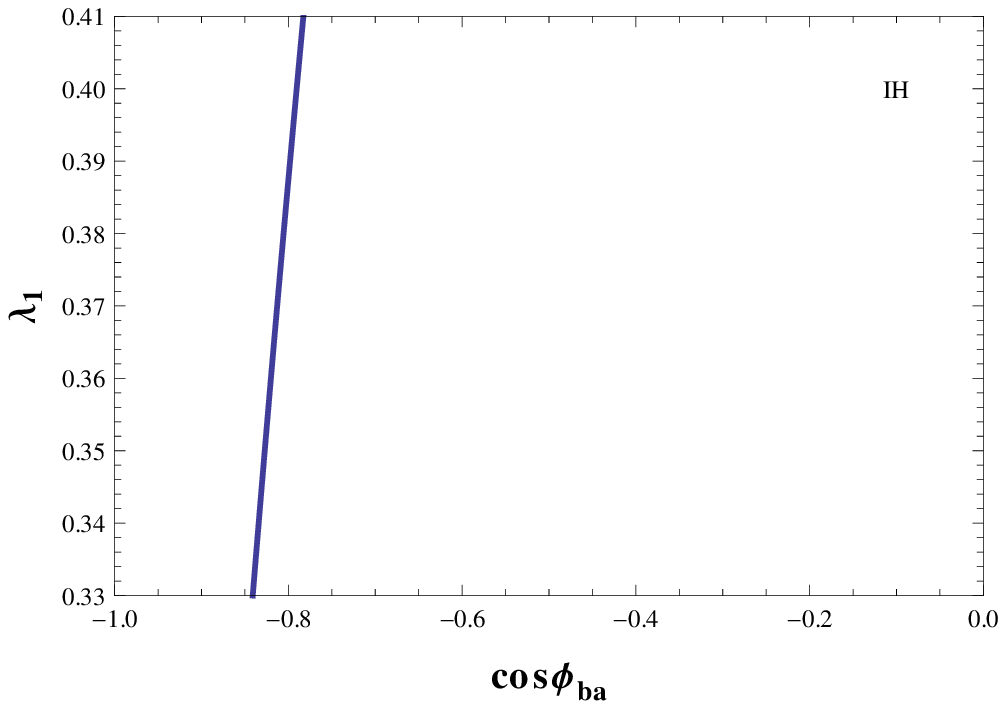}
\caption{{\small $\cos\phi_{ba}$ vs $\lambda_1$ for $\lambda_2=1$ NH and 
                 $\lambda_2=2.1$ for IH.}}
\label{csa1}
\end{figure}
This is shown in Fig.\ref{csa1}. We find that $\cos\phi_{ba}$ does not vary
much with $\lambda_1$ in the specified range. Similar to the one discussed in 
section \ref{sec4}, we can also set the Majorana phases $\alpha_{21}$ and 
$\alpha_{31}$ as a function of $\lambda_1$ and finally
we plot $Y_{B}$ against $\sin^2\theta_{13}$ in Fig.\ref{ybs13} as
$\sin^2\theta_{13}$'s dependence on $\lambda_1$ is known. Note that, here also 
we have used the values Re$(x_{C,D})=0.2$ for NH and $0.05$ for IH as before. 
The maximum value of the effective neutrino mass parameter turns out to be
$|\langle{m}\rangle|\sim$ 0.01 eV for NH (0.025 eV for IH).
\begin{figure}[h]
\centering
\includegraphics[scale=0.6]{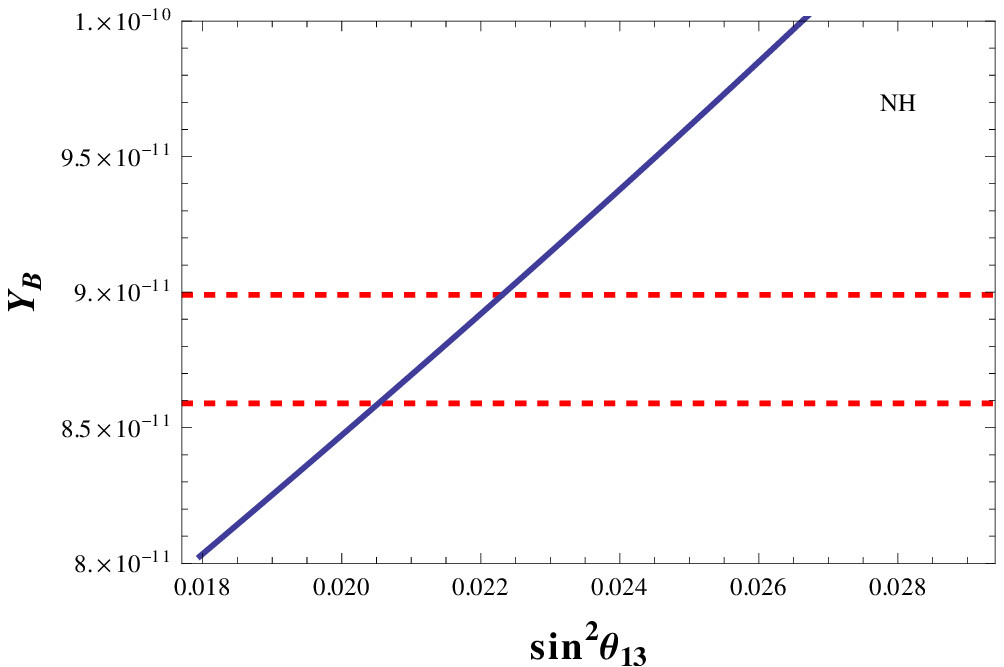}
\hspace{1cm}
\includegraphics[scale=0.6]{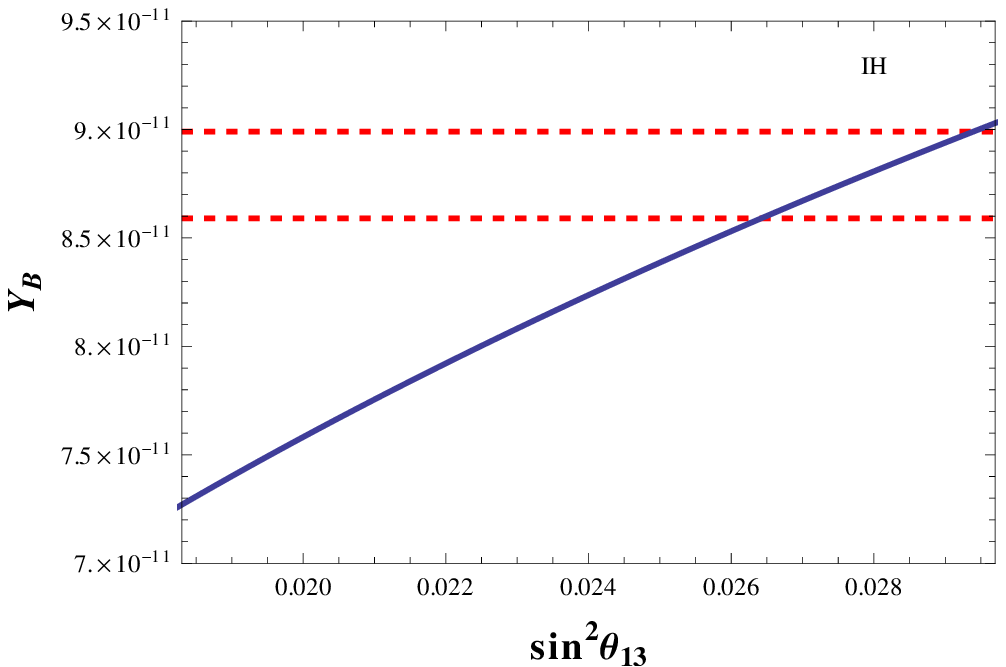}
\caption{{\small Baryon asymmetry $Y_B$  vs $\sin^2\theta_{13}$ for
                NH (left panel) and IH (right panel).  Here the region between
                horizontal dashed lines represent observed value for $Y_B$ from
                \cite{Bennett:2012zja}.}}
\label{ybs13}
\end{figure}
%%%%%%%%%%%%%%%%%%%%%%%%%%%%%%%%%%%%%%%%%%%%%%%
%%%%%%%%%%%%%%				 %%%%%%%%%%%%%%%%%%%%%%%
%%%%%%%%%%%%%	         Conclusion       %%%%%%%%%%%%%%%%%%%%%%%
%%%%%%%%%%%%%%				  %%%%%%%%%%%%%%%%%%%%%%
%%%%%%%%%%%%%%%%%%%%%%%%%%%%%%%%%%%%%%%%%%%%%%%
\section{Conclusion}\label{conc}
In this work, we have studied the generation of nonzero $\theta_{13}$ in a
$A_4$ symmetric framework. For this, we have extended the particle content
of the AF model by adding one flavon, $\xi'$. In doing so, we consider the
generation of light neutrino masses and mixing through the type-I see-saw
mechanism. The addition of $\xi'$ leads to a deformed structure for the right 
handed neutrino mass matrix as compared to the one obtained in case of 
tri-bimaximal mixing pattern. The explicit structure of the right handed 
neutrino mass matrix as well as the neutrino Yukawa matrices dictated by the 
flavor symmetry imposed ($A_4 \times Z_3$), helps in studying the mixing angles 
involved in the $U_{PMNS}$ matrix. We find that our framework can reproduce all 
the mixing angles consistent with recent experimental findings for a restricted 
range of parameter space for $\lambda_1$ involved in the theory. We find a 
modified sum rule for this particular set-up. Also the effective neutrino mass 
parameter $|\langle{m}\rangle|$ is studied. Since the 
structure of right handed neutrino sector is known, it also opens up the
possibility to study leptogenesis in this framework and particularly the
involvement of Majorana phases in the setup can be utilized.  Following
\cite{Hagedorn:2009jy}, we then study the Majorana phases $\alpha_{21}$ and
$\alpha_{31}$ involved in $U_{PMNS}$ and their dependence on parameter
$\lambda_2$, while keeping $\lambda_1$ fixed at a value that could reproduce the
best fit value of $\sin^2\theta_{13}$. This is done while constraints on
neutrino parameters like the ratio of $\Delta m^2_{21}$ and $\Delta m^2_{31}$ is
considered in conjugation with the sum rule obtained. It is known that this sort
of model will not generate lepton asymmetry due to the special form of neutrino
Yukawa matrix involved. The same conclusion holds here also and we need to
consider the next-to-leading order effect to the neutrino Yukawa sector in
order to realize nonzero lepton asymmetry. We have calculated the
next-to-leading order terms in our setup and their involvement in the expression
for the CP-asymmetry parameter $\epsilon_i$. Then we have shown that within `one
flavor approximation', our setup is able to generate sufficient amount of
lepton asymmetry through the decay of the right handed neutrinos (and
sneutrinos) without assigning large values to the parameters involved. In
obtaining this result, we use the information obtained on the Majorana phases
$\alpha_{21}, \alpha_{31}$ as function of the parameters involved.  As the
baryon asymmetry can be linked with the generated lepton asymmetry finally we
have studied the variation of baryon asymmetry parameter $Y_B$ with $\lambda_2$.
The effect of having nonzero $\theta_{13}$ is also studied.

It can also be noted that the framework restricts the RH neutrino masses in 
a narrow range between $(1+\tan^2\beta)10^{12}$ GeV and $10^{14}$ GeV as 
evident from the discussion below Eq.(\ref{eta}). This in turn can be used 
to estimate the scales involved in the theory. With our consideration that all
the vevs of the new scalars involved in the set-up to be of similar order
of magnitude, $v$, the RH neutrino masses are of order $M_i \sim 2xv$ 
as seen from Eq.(\ref{M1real}-\ref{M3real}). With coupling constants 
$x\sim\mathcal{O}(1)$, it further tells that $v$ is of order  
$10^{13}$ GeV with $\tan\beta\sim3$. Therefore the new flavons (whose masses are 
proportional to $v$ as seen from Eq.(\ref{wd})) are found to be as heavy as 
RH neutrinos, while the couplings involved are considered to be of order 1.
So although the RH neutrinos have other interactions with the new scalars of
the set-up (from Eq.(\ref{wnu})), its decay mode is essentially dominated by
the Yukawa interactions with the lepton and higgs doublets only. This justifies 
our consideration of employing Eq.(\ref{epsi}) which is the standard expression
of leptogenesis for the decay of RH neutrinos through Yukawa interaction. Now, in order to 
produce correct amount of lepton asymmetry, we require to have 
$\kappa=\frac{v}{\Lambda}$ to be of order $10^{-2}$. This value is also
consistent with the tau lepton mass as appeared in Eq.(\ref{cl}) with the
coupling $y_{\tau} \sim \mathcal{O}(1)$. This sets the typical value of 
 the cut-off scale $\Lambda$ to be $10^{15}$ GeV. The close proximity of
 $\Lambda$ with the grand unification scale turns out to be an intriguing 
 feature of the model. 
%%%%%%%%%%%%%%%%%%%%%%%%%%%%%%%%%%%%%%%%%%%%%%%
%%%%%%%%%%%%%%				 %%%%%%%%%%%%%%%%%%%%%%%
%%%%%%%%%%%%%	         Appendix       %%%%%%%%%%%%%%%%%%%%%%%
%%%%%%%%%%%%%%				  %%%%%%%%%%%%%%%%%%%%%%
%%%%%%%%%%%%%%%%%%%%%%%%%%%%%%%%%%%%%%%%%%%%%%%
\appendix
\numberwithin{equation}{section}
\section*{Appendix}
\section{$A_{4}$ Multiplication Rules:}\label{apa}
%\lipsum[2]
%\subsection{$A_{4}$ Multiplication Rules:}\label{apa}
$A_{4}$ is discrete group of even permutation of four objects\footnote[5]{For
a detailed discussion on $A_4$, see \cite{Altarelli:2010gt}.}. It has three 
inequivalent one-dimensional representation $1,1^{\prime},1^{\prime\prime}$ and
a irreducible three dimensional representation 3. Product of the singlets and
triplets are given by 
\begin{eqnarray}
1\otimes{1}&=&1,\nonumber\\
1'\otimes1'&=&1'',\nonumber\\
1'\otimes1''&=&1,\nonumber\\
1''\otimes1''&=&1', \& \nonumber\\
3\otimes3&=&1\oplus1'\oplus1''\oplus{3_{A}}\oplus{3_{S}}
\end{eqnarray}

where subscripts $A$ and $S$ stands for ``asymmetric'' and ``symmetric''
 respectively. If we have two triplets $(a_{1},a_{2},a_{3})$ and
$(b_{1},b_{2},b_{3})$, their products
are given by

\begin{eqnarray}
1&\sim&a_{1}b_{1}+a_{2}b_{3}+a_{3}b_{2},\nonumber\\
1'&\sim&a_{3}b_{3}+a_{1}b_{2}+a_{2}b_{1},\nonumber\\
1''&\sim&a_{2}b_{2}+a_{3}b_{1}+a_{1}b_{3},\nonumber\\
3_{S}&\sim&\begin{bmatrix}\nonumber
      2a_{1}b_{1}-a_{2}b_{3}-a_{3}b_{2}\\
       2a_{3}b_{3}-a_{1}b_{2}-a_{2}b_{1}\\
       2a_{2}b_{2}-a_{1}b_{3}-a_{3}b_{1}
        \end{bmatrix},\nonumber\\
3_{A}&\sim&\begin{bmatrix}
      a_{2}b_{3}-a_{3}b_{2}\\
      a_{1}b_{2}-a_{2}b_{1}\\
       a_{3}b_{1}-a_{1}b_{3}
        \end{bmatrix}.
\end{eqnarray}

\section{$A_{4}$ Vaccum Alignments:}\label{apb}
%\lipsum[3] 

%\subsection{$A_{4}$ Vaccum Alignments:}\label{apb}
In our model driving part of the LO superpotential, invariant under 
$A_{4}\times{Z_{3}}$ with $R=2$,  can be written as 
\begin{equation}\label{wd}
 w_{d}=M(\phi_{0}^{T}\phi_{T})+g(\phi_{0}^{T}\phi_{T}\phi_{T})+\phi^{S}_{0}(g_{1
}\phi_{S}\phi_{S}+g_{2}\phi_{S}\xi+g_{3}\phi_
{S}\xi')+\xi_{0}(g_{4}\phi_{S}\phi_{S}+g_{5}\xi\xi).\\
\end{equation}
Equations which give vacuum structure of $\phi_{T}$ are given by: 
\begin{eqnarray}
\frac{\partial w}{\partial
\phi^T_{01}}&=&M\phi_{T1}+\frac{2g}{3}\left(\phi^{2}_{T1}-\phi_{T2}\phi_{T3}
\right)=0, \nonumber\\
\frac{\partial w}{\partial
\phi^T_{02}}&=&M\phi_{T1}+\frac{2g}{3}\left(\phi^{2}_{T2}-\phi_{T1}\phi_{T3}
\right)=0, \nonumber\\
\frac{\partial w}{\partial
\phi^T_{03}}&=&M\phi_{T1}+\frac{2g}{3}\left(\phi^{2}_{T3}-\phi_{T1}\phi_{T2}
\right)=0. \\
\nonumber\end{eqnarray}
Solution of these equations can be given by:
$\langle\phi_{T}\rangle=(v_{T},0,0)$ where $v_{T}=-\frac{3M}{2g}$. Again,
equations responsible for vacuum alignments of $\phi_{S}$, $\xi$ and $\xi'$
are: 
\begin{eqnarray}
\frac{\partial w}{\partial \phi^S_{01}}&=&
\frac{2g_{1}}{3}\left(\phi^{2}_{S1}-\phi_{S2}\phi_{S3}\right)+g_{2}\xi\phi_{S1}
+g_{3}\xi'\phi_{S3}=0 \nonumber\\
\frac{\partial w}{\partial \phi^S_{02}}&=&
\frac{2g_{1}}{3}\left(\phi^{2}_{S2}-\phi_{S1}\phi_{S3}\right)+g_{2}\xi\phi_{S3}
+g_{3}\xi'\phi_{S2}=0\nonumber\\
\frac{\partial w}{\partial \phi^S_{03}}&=&
\frac{2g_{1}}{3}\left(\phi^{2}_{S3}-\phi_{S1}\phi_{S2}\right)
+g_{2}\xi\phi_{S2}+g_{3}\xi'\phi_{S1}=0\nonumber\\
\frac{\partial w}{\partial \xi_{0}}&=& g_{4}(\phi^{2}_{S1}
+2\phi_{S2}\phi_{S3})+g_{5}\xi\xi=0\\
\nonumber\end{eqnarray}
From these equations we obtain 
$\langle\phi_{S}\rangle=(v_{S},v_{S},v_{S}),  \langle\xi\rangle=u$ and
$\langle\xi'\rangle=u'\neq 0$ with $v_{s}^{2}=\frac{-g_{5}u^{2}}{3g_4}$ 
and $u'=\frac{-g_{2}u}{g_3}$. 
Note that NLO correction terms with $1/\Lambda$ suppression involving 
$\xi'$ in the superpotential $w_d$ are absent and so the vevs of the flavon 
fields remain unchanged.

\end{document}